# Are we facing a reproducibility crises in materials synthesis? A systematic review of Turkevich AuNP synthesis and CVD $MoS_2$ growth

*Julia S. Correa,*[1‡] *Leandro V. Silva,*[1‡] *Nichollas G. G. Silva,*[1‡] *Cesar Raitz,*[1‡] *Alex S. Lima,*[1] *Daniel Grasseschi*[1*]

[1]Department of Inorganic Chemistry, Chemistry Institute – Federal University of Rio de Janeiro Av. Pedro Calmon, 550 – CidadeUniversitária, 21941-901, Rio de Janeiro, Brazil.

‡ Equal contributions.

*Corresponding Author: dgrasseschi@iq.ufrj.br

**ABSTRACT**

Reproducibility remains a major challenge in materials synthesis, particularly for nanomaterials whose properties are highly sensitive to experimental conditions. Here, we present a systematic review and meta-analysis evaluating the reproducibility of two widely used synthesis routes: the Turkevich method for gold nanoparticles (AuNPs) and the chemical vapor deposition (CVD) growth of $MoS_2$. An adapted PRISMA-based protocol combined with a modified SPIDER framework was applied to assess methodological transparency, parameter reporting, and experimental consistency across the literature. More than 1,300 articles for each case study were retrieved from Scopus and Web of Science and systematically screened using structured checklists and a Python-based text classification algorithm validated against independent human reviewers. Despite the extensive literature and the widespread perception of these methods as reproducible, only a small fraction of studies rigorously addressed synthesis reproducibility. Critical experimental parameters were frequently underreported, and statistical analyses were rarely included, limiting inter-laboratory comparability and reproducibility. These findings demonstrate the value of systematic reviews and meta-analyses as tools for identifying reproducibility gaps and guiding the development of more transparent and reliable synthesis protocols in materials science.

## INTRODUCTION

Reproducibility is a fundamental principle of the scientific method, underpinning the credibility and cumulative nature of scientific knowledge. In materials science, where experimental systems are often highly sensitive to preparation methods, environmental conditions, and subtle procedural details, ensuring reproducibility is both particularly challenging and essential. The ability to replicate findings across laboratories not only validates results but also builds a reliable foundation for technological development and innovation.

The importance of reproducibility has gained increased attention across disciplines, driven by evidence that many published results cannot be independently verified. A survey conducted by Nature in 2016 involving 1,576 researchers revealed that more than 70% had failed to reproduce another scientist's experiments, and over half had failed to reproduce their own work.[1] This crisis has prompted initiatives such as the Reproducibility Project and the Brazilian Reproducibility Initiative, which systematically attempt to replicate published experiments in psychology, cancer biology, and biomedical research.[2] However, while several initiatives in materials science have focused on improving reproducibility in computational methods,[3] there remains a notable lack of systematic studies addressing reproducibility in materials synthesis.

In materials science, reproducibility issues are particularly pronounced in areas involving nanomaterials and surface chemistry, such as the synthesis of gold nanoparticles or bidimensional materials via bottom-up techniques. As highlighted by Bøggild (2023), large-scale nanomaterials production remains limited not only by cost but by reproducibility and consistency in characterization methods.[4,5] Researchers frequently report difficulties replicating published transfer processes, even when following protocols precisely.[1] This issue is compounded by the tendency in the literature to describe what was done without providing sufficient guidance on how to achieve similar results, what Bøggild describes as a lack of "intersubjectivity."

Given the time- and resource-intensive nature of experimentally assessing the reproducibility of synthesis methods, alternative approaches must be proposed and adopted to help identify knowledge gaps in the field. In this context, systematic reviews, particularly when accompanied by data meta-analyses of existing literature, offer a promising and efficient solution.[6–8]

Systematic reviews, which originated in the biomedical sciences in the late 1970s, aim to synthesize findings from multiple primary studies addressing a specific research question. Unlike traditional narrative literature reviews, systematic reviews are characterized by their methodological rigor, transparency, and reproducibility.[7,9,10] They follow predefined protocols for literature search, selection, data extraction, and analysis. Systematic reviews can be either qualitative or quantitative, with the latter often incorporating meta-analysis to statistically combine data from independent studies.[7]

Meta-analysis enables researchers to aggregate results from multiple papers that share common experimental variables, allowing for a broader statistical evaluation of trends and inconsistencies across studies. This approach can help uncover patterns not evident in individual studies and offers greater statistical power. Despite its demonstrated utility, systematic reviews and meta-analyses remain underutilized in many scientific disciplines, including materials science.[11–13]

Numerous methodological frameworks have been developed to guide systematic review practices, most notably in the biomedical field. Protocols such as Cochrane,[14] PRISMA (Preferred Reporting Items for Systematic Reviews and Meta-Analyses)[6], and PROSPERO (an international database of prospectively registered systematic reviews)[15] are widely regarded as gold standards for conducting and reporting systematic reviews. However, these guidelines are primarily tailored to clinical and health-related research and often require adaptation when applied to other domains, such as chemical synthesis or materials science.

Adapting these frameworks to suit the specificity of experimental research in materials science, such as variability in reaction conditions, measurement techniques, and data reporting practices, is essential for leveraging the full potential of systematic reviews in this field.

With this in mind, the present systematic review and meta-analysis aims to bridge methodological gaps between disciplines by adapting well-established protocols, particularly the SPIDER (Sample, Phenomenon of Interest, Design, Evaluation, Research type) framework,[16] to the field of materials synthesis. Specifically, we apply a modified version of PRISMA to assess the reproducibility of AuNP synthesis using the Turkevich method and CVD growth of $MoS_2$, a widely cited yet variably reported synthesis methods in the literature. These systems were selected because they represent foundational and widely studied processes in nanomaterials research, both central to modern materials science. Gold nanoparticles serve as a benchmark system for colloidal synthesis, exhibiting tunable optical, catalytic, and electronic properties that are highly sensitive to subtle variations in experimental conditions.[17–31] The Turkevich method, in particular, is considered a "standard" synthesis route, yet its reproducibility remains under debate due to inconsistencies in methodological reporting and parameter control across studies.[17,20,31–35]

Similarly, $MoS_2$ is a prototypical two-dimensional material whose CVD growth involves complex interdependencies between precursor composition, substrate choice, temperature, and gas flow dynamics.[36–45] As one of the most investigated 2D material, its reproducible synthesis is critical for scalable production and integration into optoelectronic and catalytic applications. Together, these two cases provide complementary perspectives on reproducibility, one from solution-phase synthesis and another from vapor-phase growth, allowing us to evaluate how methodological transparency and parameter reporting impact reproducibility across different synthesis paradigms in materials science.

A preliminary search using the keyword "AuNP", "Turkevich", and "reproducibility" identified 114 papers that mentioned terms related to statistical analysis or reproducibility. However, only 20 of these references were directly related to the synthesis process itself through approaches such as batch-to-batch comparison, factorial experimental design, or parameter optimization.[46–65] The remaining studies referred to the reproducibility of nanoparticle applications rather than their synthesis. Notably, 7 out of the 20 synthesis-focused articles incorporated reproducibility analysis as part of efforts to transition from classical batch synthesis to continuous flow systems.[49,51,55,58,59,64,65]

Searching for the terms **"$MoS_2$"** and **"CVD"**, we identified **173 publications** containing keywords associated with reproducibility, systematic evaluation, or statistical analysis. Nevertheless, in most cases, *reproducibility* referred to the performance of fabricated devices or specific material applications rather than to the reproducibility of the CVD synthesis itself. Of the studies discussing the growth process, **38%** focused on method optimization, generally relying on one-factor-at-a-time parameter variation instead of statistically designed reproducibility studies. Remarkably, **only one publication** reported a comprehensive assessment of the reproducibility of the CVD growth process.[66]

Despite the widespread perception of both methods as a reliable and reproducible routes, there is a limited number of systematic studies that rigorously evaluate this assumption. Several factors may contribute to this gap, including the time- and resource-intensive nature of full factorial experimental designs, and the large number of replicates required to validate new synthesis protocols under statistically meaningful conditions.

These limitations are also reflected in how experimental methodologies are reported. Incomplete descriptions, along with the lack of standardized reporting protocols, frequently hinder the accurate replication of experiments across research groups, ultimately compromising reproducibility in the field. This leads to a critical question: is the synthesis process inherently irreproducible due to the multitude of variables that must be controlled, or is the apparent irreproducibility a consequence of inadequate methodological reporting?

## METHODOLOGY

This study followed established guidelines for conducting systematic reviews and meta-analyses, with methodological adaptations tailored to the field of materials synthesis.[14–16] To ensure transparency and reproducibility, a systematic review protocol was developed and documented detailing all steps, inclusion/exclusion criteria, and decision-making strategies.

### 1. Review Question Formulation

The guiding research question was formulated using an adapted version of the SPIDER framework, originally developed for qualitative evidence synthesis, to accommodate the specificities of experimental studies in materials science.[16] The SPIDER elements were interpreted as follows:

- **Sample (S):** Articles reporting the synthesis of gold nanoparticles via the Turkevich method or CVD growth of $MoS_2$.
- **Phenomenon of Interest (PI):** Reproducibility and reporting of experimental conditions affecting the material morphology.
- **Design (D):** Primary experimental studies involving batch synthesis of AuNPs or atmospheric pressure CVD growth of $MoS_2$.
- **Evaluation (E):** Clarity and completeness of methodological reporting, consistency in

material morphology (size, shape, thickness) outcomes.

- **Research type (R)**: Quantitative primary research articles.

## 2. Database Search Strategy

A comprehensive search strategy was developed and tested using 15 reference articles (often called as sentinel articles) widely cited and recognized as reference studies in the field of AuNP synthesis[17–31] and 18 in the field of CVD growth of $MoS_2$.[67–84] These were used as a benchmark to ensure that the proposed search strings were effective in retrieving core literature. Searches were conducted across two major databases: Scopus and Web of Science. The search period was limited to publications from 1999 onward for the AuNP synthesis and from 2010 for $MoS_2$ CVD growth, and the final search strings were designed to maximize retrieval of relevant articles while ensuring precision. The used strings for both searches are shown in the support information.

## 3. Inclusion and Exclusion Criteria

After the initial retrieval of records, all articles were screened according to predefined inclusion and exclusion criteria to ensure methodological consistency and relevance to the research objectives. Only studies written in English and classified as primary research articles focusing specifically on either the synthesis of AuNP via the Turkevich method or the CVD growth of $MoS_2$ were considered eligible. To qualify for inclusion, studies were required to report quantitative synthesis outcomes, such as nanoparticle size and size distribution in the case of AuNPs, or growth characteristics, such as film thickness, grain size, or layer number for $MoS_2$. Conversely, review papers, conference proceedings, and editorials were excluded from the analysis. Articles published in journals identified as predatory according to the PredaQualis[85] and Predatory Journals databases[86] were also removed to ensure the scientific integrity of the dataset. Finally, duplicate records retrieved from multiple databases were systematically identified and eliminated prior to screening, guaranteeing that only unique and relevant studies proceeded to the subsequent abstract evaluation phase.

## 4. Abstract Screening

To ensure the inclusion of only studies in the scope of our systematic review, abstracts were evaluated using a structured checklist adapted from the STROBE (Strengthening the Reporting of Observational Studies in Epidemiology) guidelines. The checklist was tailored to assess reporting clarity, methodological transparency, and alignment with the review's objectives, specifically focusing on the Turkevich synthesis of gold nanoparticles and the CVD growth of $MoS_2$. Each item received a weighted score, with eliminatory criteria applied to confirm the correct synthesis method. Each question was answered with a binary "Yes" or "No," and cumulative scores determined inclusion. Abstracts scoring ≥75% and meeting all eliminatory criteria proceeded to full-text analysis. **Supplementary Notes 2 and 6** in the support information summarizes the cheklist questions and the weights for the AuNP and $MoS_2$ abstract screening, respectively.

Two independent reviewers conducted the screening, and inter-rater reliability was quantified using Cohen's Kappa coefficient (κ), ensuring objectivity and consistency. To complement the manual process, a Python-based keyword classification algorithm was developed to automate screening. The algorithm matched abstracts against predefined keyword sets corresponding to each checklist item. Keyword lists were iteratively refined using a training set of 140 articles until achieving agreement with human evaluations higher than 70%, calculated by the κ coefficient. After training the algorithm was applied in the entire database. The final algorithm working principal along with the keyword lists are available in the **Supplementary Note 2 and 6** and **Tables S1 and S6** Supporting Information and the complete code through the Open Science Framework.[87]

## 5. Methodological Evaluation

Full-text articles that passed the abstract screening underwent a detailed methodological assessment focused on identifying whether key experimental variables were adequately reported. This stage aimed to evaluate the feasibility of the synthesis procedures by verifying the presence or absence of critical parameters known to affect reaction mechanisms and final material characteristics.

A structured methodological checklist was developed, inspired by an adapted version of the STROBE framework, and tailored to the specific requirements of materials synthesis. Each question addressed essential aspects of experimental transparency and reproducibility, and responses were binary ("Yes"/"No"). Studies achieving ≥ 75% compliance were considered methodologically sound and included in the subsequent meta-analysis.

For gold nanoparticle synthesis via the Turkevich method, the checklist included parameters such as precursor concentration, reaction temperature, stirring rate, reagent addition order, and solution pH (see **Table S2 and S4** in the Supporting Information). For $MoS_2$ growth by CVD, variables such as sulfur and $MoO_3$ precursor masses, temperature ramp, growth time, quartz tube diameter, and precursor distance were evaluated (see **Table S7 and S8** in the Supporting Information). Failure to report any of these parameters hinders reproducibility and comparability across studies. The checklist also assessed procedural clarity and experimental robustness, including glassware cleaning, number of replicates, and control of environmental conditions. Studies focusing solely on material applications or using commercially available samples, rather than describing synthesis procedures, were excluded at this stage. All the methodological evaluation was carried only by human evaluators.

## 6. Data Extraction and Meta-Analysis

From the final set of included studies, detailed synthesis parameters and reported materials' characteristics were extracted and tabulated. This dataset formed the basis for a quantitative meta-analysis, conducted using the METAFOR R programming environment. The meta-analysis aims to evaluate whether similar experimental conditions across independent studies yield statistically consistent results, and to identify which factors most critically influence reproducibility.

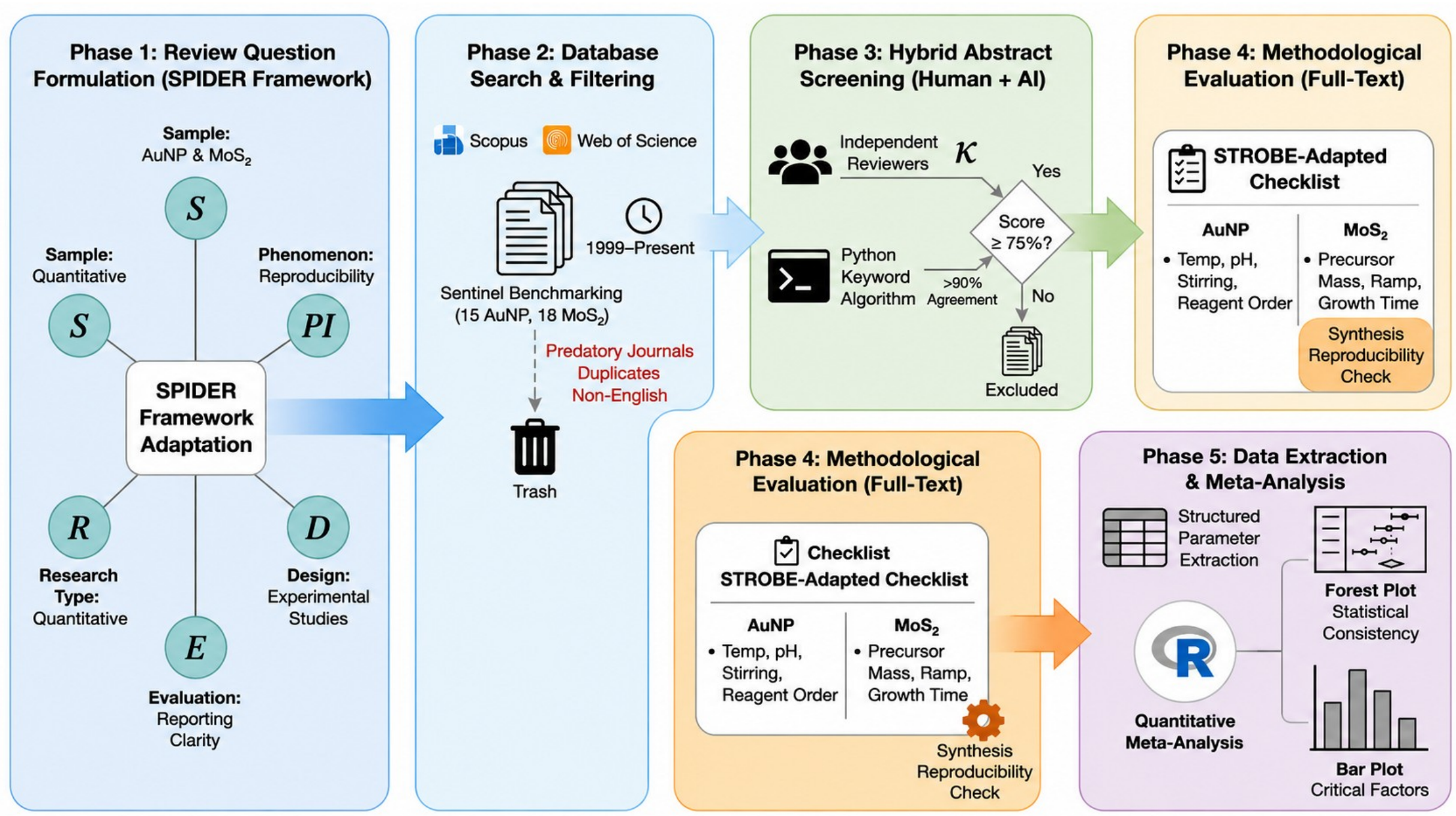

**Figure 1:** Workflow of the systematic review methodology following a SPIDER-based framework. Phase 1 defines the research question and inclusion criteria using the SPIDER strategy. Phase 2 involves database search and filtering (Scopus and Web of Science), including removal of duplicates, non-English articles, and low-quality sources. Phase 3 combines human and AI-assisted hybrid abstract screening based on predefined agreement and scoring criteria. Phase 4 consists of full-text methodological evaluation using a STROBE-adapted checklist, with emphasis on synthesis reproducibility for AuNP and $MoS_2$ studies. Phase 5 includes structured data extraction and quantitative meta-analysis, integrating statistical consistency assessment and comparative evaluation of reported parameters.

## RESULTS AND DISCUSSION

Using the adapted STROBE-based framework, we systematically identified, screened, and evaluated published studies to determine how methodological reporting affects reproducibility and comparability across research groups. Our objective is to critically evaluate whether published studies that report similar experimental conditions yield reproducible outcomes in terms of particle size, distribution, and morphology. In doing so, we seek to determine whether the perceived reproducibility of the method is supported by consistent empirical evidence, or if instead, reproducibility is undermined by inadequate methodological reporting. Through this adapted systematic approach, we aim to identify which experimental parameters are most frequently reported, which are often omitted, and which appear to be most critical in enabling the replication of results across different research groups.

## Gold Nanoparticles reproducibility systematic review

AuNPs have long attracted scientific attention due to their distinctive size-dependent optical,[88,89] electronic,[90] and surface properties, which enable applications in catalysis,[91,92] biosensing,[93] drug delivery,[94] and nanoelectronics.[95–97] Although their optical behavior was first observed by Faraday in 1856,[98] a systematic synthesis route was only established in 1951, when Turkevich introduced the citrate reduction method using sodium citrate as both reducing and stabilizing agent (**Figure 2A**).[99–103] Turkevich work laid the foundation for understanding how reaction parameters such as temperature, reactant concentration, and precursors concentration affect nucleation and particle growth (**Figure 2B**). In 1973, Frens refined the approach by adjusting the citrate-to-gold ratio, enabling control over nanoparticle size and uniformity.[104] Later, Zukoski and collaborators (1994) proposed a revised mechanism involving particle aggregation and disaggregation, emphasizing the sensitivity of the synthesis to concentration and reagent ratios.[105] Subsequent studies have shown that additional factor, including pH, stirring rate,[17] reagent addition order,[106] and heating profile[17] further influence particle morphology and stability (**Figure 2B**).[105,107–114] Nowadays it is very well established that the precise control of the experimental parameters are crucial to tune the nanoparticles size, shape and distribution and consequently manipulate their optical properties (**Figure 2C**) for different applications.

Despite extensive research on gold nanoparticle synthesis, the Turkevich and Frens-modified Turkevich methods remain the most widely used and are generally regarded as reproducible and reliable. However, few studies have systematically isolated and quantified the influence of individual parameters on reproducibility using statistically robust designs. Consequently, questions persist about the true reproducibility of these synthesis protocols across laboratories and experimental conditions.

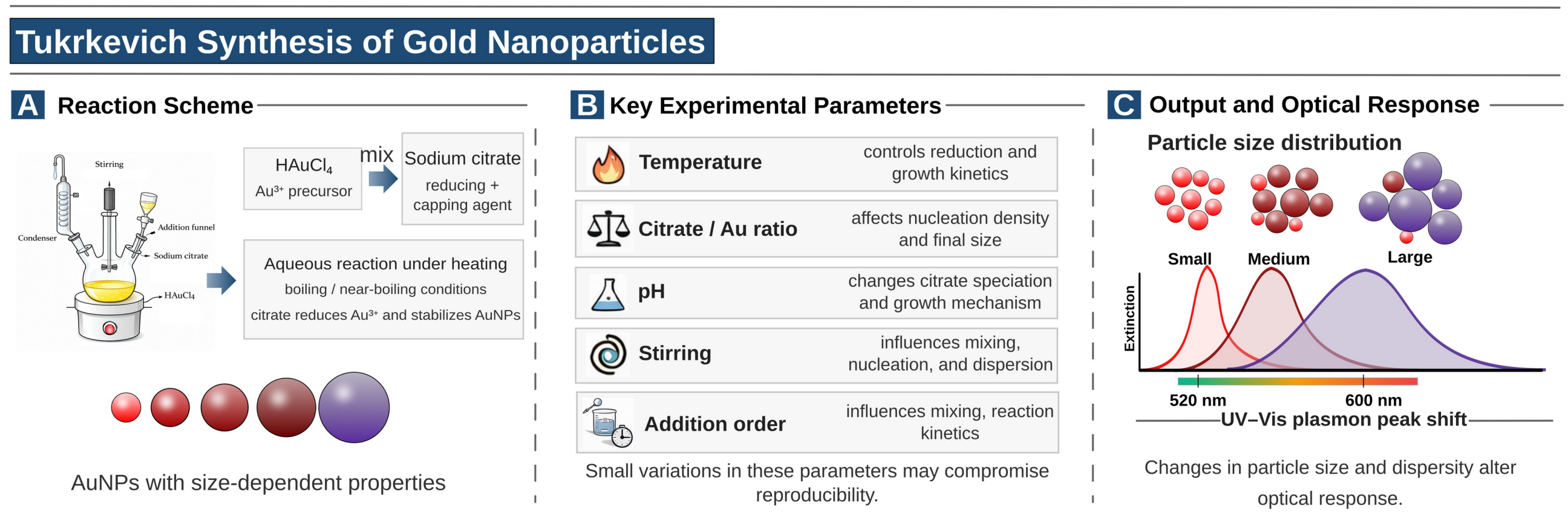

**Figure 2: Schematic representation of the Turkevich synthesis of AuNPs**.
The citrate-mediated reduction of $HAuCl_4$ in boiling aqueous solution is illustrated, where citrate acts as both reducing and stabilizing agent. The reaction proceeds through nucleation followed by particle growth, with citrate providing electrostatic stabilization of the formed AuNPs. Key experimental parameters—such as precursor concentration, citrate-to-gold ratio, temperature, stirring, and pH—govern nanoparticle size and dispersity. The formation of AuNPs is accompanied by a characteristic color change due to localized surface plasmon resonance (LSPR).

### AuNP Studies Abstract Screening.

The search strategy was refined using sentinel papers[17-31] to optimize retrieval in the Scopus and Web of Science databases. The final query yielded 1,233 articles from Scopus and 1,317 from Web of Science for the 1999–2024 period. After removing duplicates, 1,494 unique articles were retained. Following exclusions based on language, publication type, and journal quality (according to the PredaQualis and Pedratory Journals database)[85,86], 1,303 articles remained for abstract screening. An additional four were excluded due to overlapping content (same title, authors and abstract), resulting in a final dataset of 1,299 articles entering the first screening phase (**Figure 3A**).

First, 10% of the total number of articles were randomly selected for a validation step in which two independent reviewers and the developed Python-based algorithm applied the abstract screening checklist. Each reviewer and the algorithm completed the checklist by answering every question and calculating a final score that determined whether the article was categorized as approved or reproved. The level of agreement among reviewers and between human and algorithmic evaluations was measured using κ coefficient. Discrepancies were discussed collaboratively, leading to adjustments in the inclusion/exclusion criteria and iterative refinement of the algorithm's keyword list. This validation process was repeated until a concordance level ≥ 75% was achieved, ensuring high reliability in automated classification. The **Supplementary Note 2** and **3** in the support information discuss the details of the algorithm development and the validation step. Once validated, the finalized algorithm was applied to the entire AuNP article database, with 1299 articles, for large-scale abstract screening. A total of 466 articles were approved in this step, representing a total of 36% (**Figure 3B**).

**Table S2** in the Supporting Information summarizes the checklist questions and their respective weights used during the abstract screening. While, **Table S3** show the abstract checklist results. In this step, the questions target key elements considered essential for a well-structured title and abstract. To be approved for the methodology review, each paper needed to achieve a score ≥ 75%. At this stage, the evaluation focused on general aspects such as title clarity, authorship visibility (verifying whether the corresponding author was identifiable), clarity of objectives or hypotheses, and conclusion coherence. More specific aspects related to experimental detail were also considered, including methodological adherence to the Turkevich method, the use of direct particle size measurement techniques (e.g., electron microscopies, atomic force microscopy, dynamic light scattering), the mention of indirect methods (e.g., UV–Vis spectroscopy), and whether the average particle size or size distribution was reported. Together, these criteria ensured that only abstracts demonstrating clarity, completeness, and methodological relevance for the scope our systematic review advanced to the next evaluation stage.

The question concerning methodological adherence to the Turkevich synthesis served as an exclusion criterion, ensuring that only studies explicitly employing the classical Turkevich method were included in the analysis. This step acted as a preliminary filter, allowing the selection of studies that shared a comparable experimental framework. Following this screening, 466 articles (31%) were retained for further evaluation.

The checklist also included questions related to nanoparticle size characterization, designed to assess whether the synthesis results were adequately reported in terms of size and size distribution, key indicators of product quality and reproducibility. In this context, the use of direct characterization techniques was given higher weight than indirect methods, as they provide more accurate and quantitative information about particle size. Among the analyzed papers, 55% and 28% reported on the abstract results obtained through direct and indirect methods, respectively. The complete abstract reporting profile is shown in **Figure 3C.**

Additionally, a specific question was introduced to verify whether statistical analysis was applied to interpret particle size data, as the inclusion of measures such as standard deviation, size variance, distribution histograms, synthesis replicates enhances confidence in the reproducibility of the reported synthesis. For this criterion, any explicit mention of size variance, particularly when linked to replicate syntheses, was considered valid. Nevertheless, only 21% of the abstracts included this information, suggesting a general lack of statistical rigor and limited transparency in summarizing experimental results within article abstracts.

**AuNP Methodological evaluation.**

The methodological descriptions of the 466 articles selected after abstract screening were evaluated. At this stage, 9 articles were excluded due to paywall restrictions, resulting in a final set of 457 articles for analysis. **Table S4** in the Supporting Information summarizes the checklist questions and their respective weights used to calculate the methodological score. To be considered approved, an article was required to achieve a threshold ≥ 75%. **Figure 3D** shows the workflow followed in the methodological analysis.

This evaluation focused on the completeness and clarity of key experimental details, including the step-by-step synthesis procedure, reactant concentrations and volumes, controlled physical parameters, and the overall level of methodological description. Two questions were treated as exclusionary criteria. The first concerned the reporting of gold and citrate precursor concentrations, ensuring the exclusion of studies using commercially available or donated nanoparticles; this criterion also received the highest weight (3), as it distinguishes studies that actively perform the synthesis. The second exclusionary criterion addressed the addition of other reactants, restricting the dataset to the classical Turkevich method without modifications, with a base weight of 1. Studies that failed to meet the criteria of one of the exclusionary questions received scores between 0% and 20%, as illustrated in **Figure 3E**. Notably, syntheses performed in flow reactors were considered as modified approaches and therefore excluded, ensuring that only batch-based systems were analyzed. At this stage more than 150 articles were removed from the methodological evaluation. Consequently, in addition to meeting the 75% score threshold, eligible studies were required to report in-house synthesis and adhere strictly to the classical Turkevich protocol.

None of the 458 articles analyzed achieved a perfect score, and only 19 studies reached the threshold of 75% of the points (**Figure 3E**), corresponding to 4% at this stage and 2% of the initial database. This analysis also allowed us to identify which methodological parameters were most frequently reported, providing an initial indication of where reproducibility issues may arise. These results are summarized in **Figure 3F** and **Table S5** of the Supporting Information. During the methodology evaluation process the checklist was applied sequentially, and once an article failed an exclusionary criterion, the remaining items were not evaluated. Using this approach, 351 out of 458 articles (77%) were confirmed to report in-house synthesis, and among these, 276 articles (79%) adhered to the classical Turkevich method. A notable observation is the low proportion of studies reporting and controlling solution pH, with only 7% explicitly describing this parameter (**Figure 3F)**. This is particularly surprising given that Turkevich himself highlighted the influence of pH on nanoparticle size and morphology as early as 1952, suggesting that this critical variable remains underreported despite its known importance.

Alongside pH, stirring rate and the reporting of replicates were among the least documented parameters, appearing in only 7% and 2% of the studies, respectively (**Figure 3F**). This is noteworthy given the importance of mixing conditions, whether magnetic or mechanical stirring, in controlling heat and mass transfer, which directly influence nucleation and growth processes (**Figure 2**). One interesting case concerns how synthesis temperature is reported, only 34 % of articles specified the temperature correctly, by measuring the temperature in the solution during synthesis. Many articles simply cited the boiling point of water, which can vary with the region where the procedure was performed. Surprisingly, some studies reported the set temperature shown on the hot-plate screen, ranging from 200 °C to 500 °C. This reveals a fundamental error in failing to recognize that the boiling or reflux temperature of water remains constant throughout the process. Additionally, each hot plate has a different power level, and the value shown on the screen is meaningless.

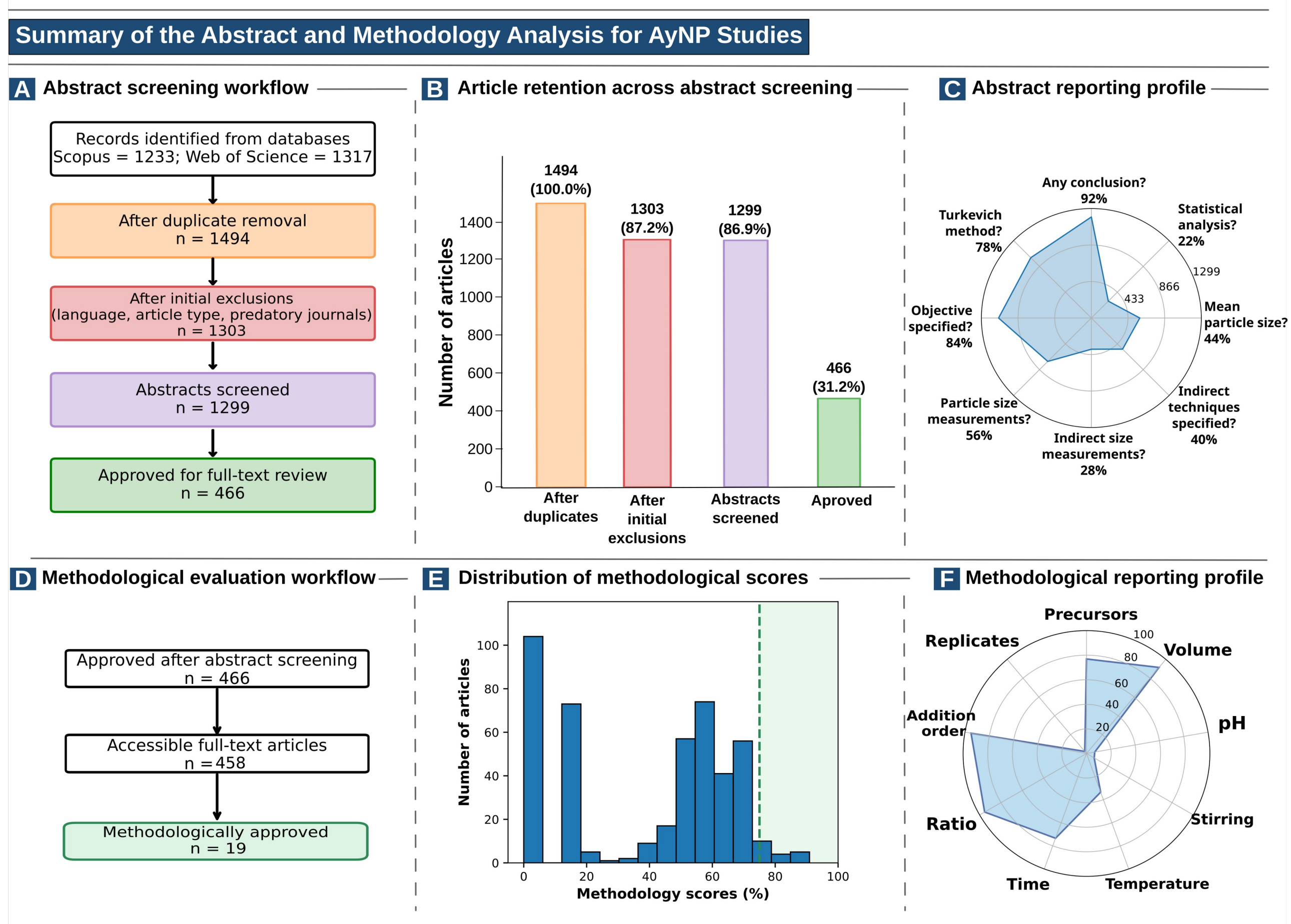

**Figure 3: Summary of the methodological evaluation of AuNP synthesis studies.**
(A) Workflow of article selection from initial database retrieval (Scopus and Web of Science) through duplicate removal, preliminary exclusions, and abstract screening, culminating in 466 articles approved for full-text analysis. (B) Article retention across screening stages, highlighting the progressive reduction in the number of studies and the relative percentage retained at each step. (C) Abstract reporting profile based on the screening checklist, illustrating the frequency with which key elements—such as synthesis method identification, characterization techniques, statistical analysis, and conclusion clarity—are reported. Together, these panels provide a comprehensive overview of the filtering process and reveal common gaps in abstract-level reporting that may impact reproducibility assessment. (D) Workflow of the methodological screening, showing the transition from 466 articles approved after abstract screening to 457 accessible full-text studies, of which only 20 met the methodological quality threshold. (E) Reporting frequency of key experimental parameters influencing AuNP synthesis, including precursor concentration, temperature, stirring, reaction time, and reagent ratios, highlighting substantial variability and frequent omissions across studies. (F) Distribution of methodological scores, illustrating the overall low compliance with reporting criteria and the limited number of studies achieving the ≥75% threshold. (D) Methodological reporting profile, summarizing the extent to which critical synthesis parameters are described, revealing systematic gaps in experimental reporting that may compromise reproducibility.

The limited reporting of replicates further highlights a declining emphasis on repeating synthesis batches under identical conditions, particularly in studies focused on applications rather than parameter optimization or kinetic control. This omission has important implications since without replicate experiments, it becomes difficult to assess the intrinsic variability of a given synthesis method, define acceptable ranges for nanoparticle size, or estimate batch-to-batch variation and statistical dispersion (e.g., standard deviation of mean size). Consequently, other researchers are left without a reliable framework to compare their results or evaluate reproducibility with statistical confidence.

These findings reinforce one of the central motivations of this systematic review, that was evaluate how effectively synthesis methodologies are described to ensure reproducibility. Even when considering only the parameters selected for meta-analysis, a substantial amount of missing information was observed, as summarized in **Figure 3F**. While some degree of incomplete reporting is expected given the large number of studies analyzed, the extent of missing data, particularly for parameters recognized as critical for over 70 years, was unexpected and may introduce bias into the meta-analysis. It is worth noting that only nine studies explicitly reported conflicts of interest, none of which were related to intellectual property associated with the synthesis method. Therefore, there appears to be no justifiable reason for the underreporting of experimental parameters.

The relatively low number of approved articles after the abstract and methodological screening raised concerns that our inclusion criteria during the abstract evaluation might be overly restrictive. To address this, the abstract screening was relaxed and all studies were the synthesis was described as the Turkevich method, citrate reduction, or closely related approaches were aproved to the methodological evaluation. The detailed classification and reasons for inclusion/exclusion are detailed in **Supplementary Note 4.2** of the Supporting Information. Adopting this broader approach increased the approval rate during the abstract screaming from 36% to 75%, resulting in a total of 978 articles.

It is important to recognize that the title and abstract are often the primary entry points through which researchers engage with a study. An abstract that lacks clarity in objectives, methodology, or key results may discourage readers from further exploration, even when the underlying work is relevant. The screening criteria applied in this study aim to capture the minimum level of information required for effective scientific communication. However, such rigor also introduces a trade-off: while it ensures consistency and transparency in study selection, it may inadvertently exclude articles with sound methodologies that are insufficiently described at the abstract level. This highlights the broader challenge of balancing strict, reproducible screening protocols with the risk of overlooking valuable contributions due to limitations in reporting practices.

Even when relaxing the inclusion criteria during the abstract screening, the same fundamental issue persists with poorly described synthesis methodologies. The most reported parameters were the reactants volume, concentration and addition order, with more than 93%. In other hand, the worst reported parameters were the solution pH (6%), stirring rate (7%), and number of replicates (2%). The reporting profile of each experimental parameter was highly similar to that observed after a more restrictive abstract screening (see **Figure S4**). This more classical approach may substantially increase the number of articles after the abstract screening, doubling the dataset for the meta-analysis stage, without necessarily improving the overall quality of the information available. One possible strategy to address this would be to also relax the criteria applied during methodological screening. However, this would result in the inclusion of a larger number of studies with incomplete or missing experimental details, requiring additional assumptions or data treatment prior to meta-analysis. Such compromises could introduce bias and ultimately weaken the reliability of the reproducibility assessment.

## Bidimensional $MoS_2$ reproducibility systematic review

Transition metal dichalcogenides (TMDs), with general formula $MX_2$ (M = transition metal, X = chalcogen), are layered materials of significant interest in two-dimensional (2D) systems. Among them, $MoS_2$ stands out as one of the most studied and technologically relevant. Its structure consists of S–Mo–S layers held together by weak van der Waals interactions, enabling exfoliation down to monolayers. A key breakthrough was the demonstration that monolayer 2H-$MoS_2$ exhibits a direct bandgap (~1.8 eV), in contrast to the indirect bandgap of its bulk form, greatly enhancing its potential for optoelectronic, catalytic, sensing, and energy applications.[115,116]

$MoS_2$ also exhibits multiple polymorphs with distinct properties, including semiconducting 2H, insulating 3R phases, and the metallic 1T phase.[117,118] Beyond thickness control, its properties can be further tuned through phase engineering, defect modulation, and doping, which strongly influence its electrical, optical, and catalytic behavior.[119,120] Consequently, achieving precise control over these structural features during synthesis remains a critical challenge for ensuring reproducibility and enabling practical applications.

Among bottom-up approaches, Chemical Vapor Deposition (CVD) stands out as a practical route for producing large-area, high-crystallinity $MoS_2$ monolayers.[121–123] In a typical process, molybdenum (e.g., $MoO_3$) and sulfur precursors are sublimated and transported to the substrate, where nucleation and growth occur (**Figure 4A**). Controlling nucleation density and growth kinetics is essential for tuning film thickness and morphology.[124] CVD-grown $MoS_2$ commonly exhibits triangular domains arising from anisotropic edge growth, although this morphology is highly sensitive to experimental conditions. Key parameters—including the Mo:S ratio, precursor temperatures, total pressure, and carrier gas flow—govern nucleation density, domain size, and flake geometry (**Figure 4B**).[125] Variations in sulfur supply, exposure time, or timing can induce transitions in shape (e.g., triangular to hexagonal), layer number, and lateral size, reflecting significant changes in growth kinetics (**Figure 4C**).[126,127] Additionally, geometrical parameters such as quartz tube diameter, precursors relative position, and substrate position influence the mass transport regime and thereby affect the growth mechanism (**Figure 4B**).[128,129]

Despite these advances, reproducibility remains a major challenge, as subtle variations in the experimental parameters can lead to significant differences in morphology, phase stability, and defect density across samples and laboratories. Although the large body of literature on CVD-grown $MoS_2$ reflects its status as the predominant method for producing high-quality 2D films, and is often implicitly regarded as reliable, systematic studies that isolate and quantitatively evaluate the influence of individual growth parameters remain scarce. Consequently, the true reproducibility of $MoS_2$ CVD synthesis across different laboratories, furnace configurations, and processing conditions is still not well established.

**CVD growth of $MoS_2$ Studies Abstract Screening**

We applied the same adapted STROBE-based systematic review framework used for the evaluation of AuNP synthesis, as described in the methodology section. A comprehensive literature search was conducted using the Scopus and Web of Science databases, considering peer-reviewed articles published between 2012 and 2024. The full search strategy is detailed in the Supporting Information. The search retrieved 1,662 records from Web of Science and 1,417 from Scopus. After removing duplicates, reviews, and articles from predatory scientific journal a pre-analysis of the title was conducted to remove works completely out of the scope, leading a total of 1,573 unique articles (**Figure 5A**). These records were then subjected to a structured screening protocol to identify studies eligible for meta-analysis. Eligible articles were those reporting the CVD synthesis of monolayer or few-layer $MoS_2$. Studies primarily focused on applications were included only when they provided a detailed and explicit description of the synthesis methodology.

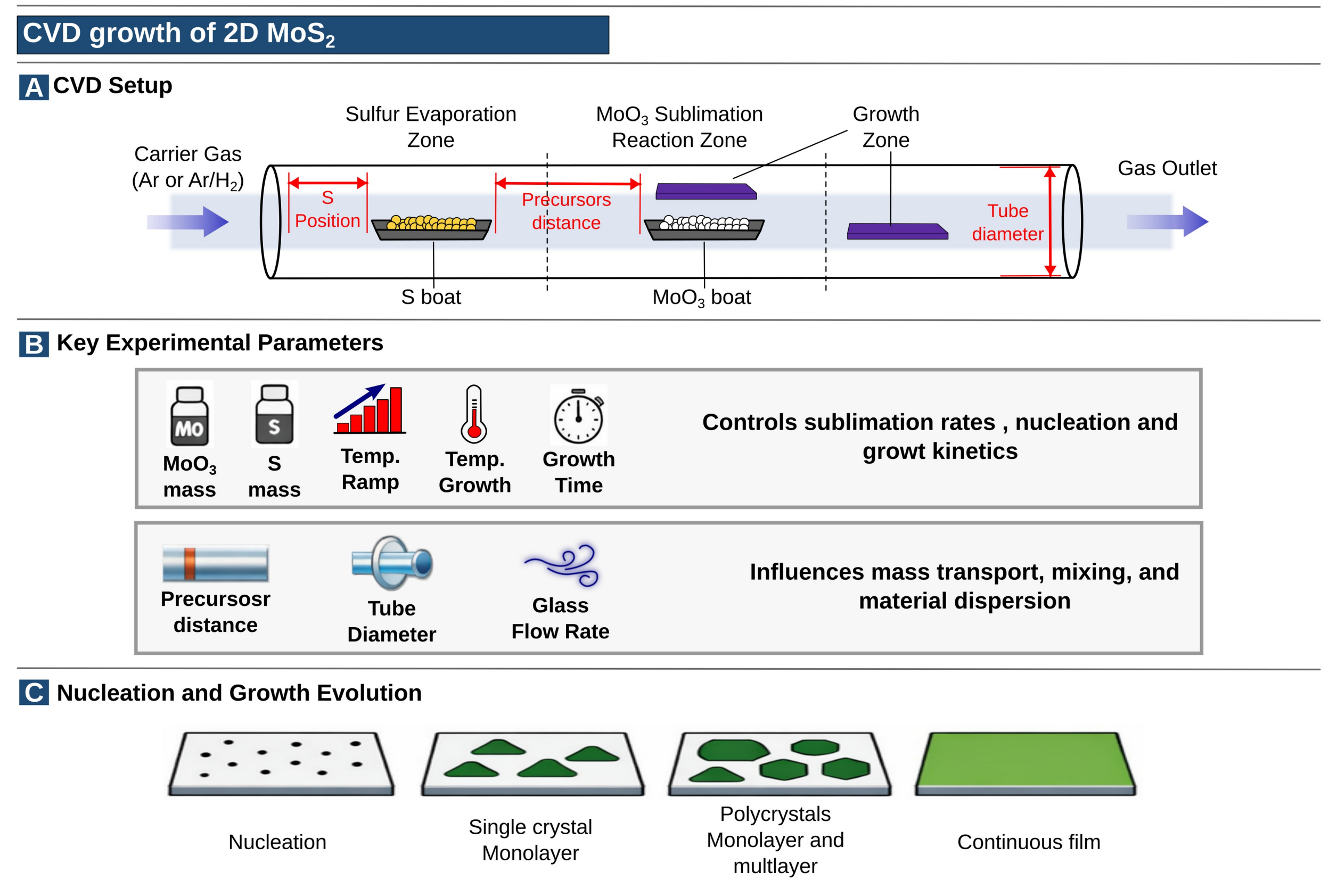

**Figure 4: Schematic representation of the key parameters governing the CVD growth of two-dimensional $MoS_2$.** (A) Typical atmospheric-pressure CVD setup illustrating the sulfur evaporation zone, $MoO_3$ sublimation/reaction zone, and growth region, highlighting critical geometrical parameters such as precursor position, precursor–substrate distance, and tube diameter. (B) Main experimental parameters affecting $MoS_2$ growth. Chemical and kinetic parameters—including precursor masses, temperature ramp, growth temperature, and growth time—primarily control precursor sublimation, nucleation, and crystal growth kinetics. In parallel, reactor geometry and transport-related parameters, such as precursor distance, tube diameter, and carrier gas flow, strongly influence fluid dynamics, mass transport, and precursor distribution within the furnace. (C) Schematic evolution of $MoS_2$ growth, from nucleation and isolated monolayer domains to polycrystalline and continuous films. Small variations in experimental parameters can significantly affect flake size, morphology, layer number, and film coverage.

Titles and abstracts were examined to determine whether the study fit the scope of the systematic review. The evaluation focused on general aspects such as title clarity, alignment between the title and the objective of the article, identification of the growth method, and clarity of conclusions. The explicit mention of 2D $MoS_2$ synthesis and the identification of “classical” or ambient pressure CVD as the growth method in the abstract were treated as exclusionary criteria, ensuring that only studies within the defined experimental framework were included. The inclusion of statistical analysis, data distribution, or reproducibility-related information was also considered with smaller weight, reflecting its importance while acknowledging that such aspects are not always fully detailed in abstracts. **Table S7** in the Supporting Information summarizes the checklist questions and their respective weights used during the abstract screening.

The search and classification algorithm previously developed for the AuNP case study was adapted and retrained using terms specifically related to $MoS_2$ CVD growth (details are discussed in **Supplementary Note 6** and **Table S6** in the Supporting Information). After validation, the algorithm selected a total of 574 articles (37% of the dataset), with an average screening score of 85%, with 109 studies achieving the maximum score, the results are summarized in **Figure 5B**. **Figure 5C** shows the reporting profile of the abstract screening checklist, and reveal that approximately 50% o the initial articles set explicitly reported the use of characterization techniques in the abstract. In contrast, just 24% mentioned parameter optimization, statistical analysis, or reproducibility assessment of the synthesis process, highlighting the limited emphasis placed on systematic evaluation of CVD growth reproducibility in the literature. The complete abstract screening results are shown in **Table S8**.

**CVD growth of $MoS_2$ Methodological Evaluation**

The experimental parameters described in the methodology section—and, when explicitly provided, in the Supporting Information—were systematically evaluated. Figures referenced in the methodology section were also considered when growth parameters were presented only graphically, such as substrate dimensions or heating ramp profiles. Three exclusionary questions were applied to ensure that the analyzed studies were within the scope of this review and described an original $MoS_2$ synthesis performed by the authors: (1) “Is the material commercial or obtained through collaboration?"; (2) "Was the material synthesized using a modified CVD methodology?"; and (3) "Does the article report the CVD growth of $MoS_2$?". These criteria ensured that only studies describing conventional CVD synthesis performed by the research group itself were included in the analysis.

Based on 26 sentinel articles, the key experimental parameters required to evaluate the reproducibility of $MoS_2$ CVD growth were identified, and weights were assigned according to their relevance to synthesis reproducibility, as detailed in **Supplementary Note 8** in the Supporting Information. The highest-weight parameters included the masses of the metallic precursor (e.g., $MoO_3$) and sulfur, carrier gas composition and flow rate, distance between precursor boats, system pressure, growth temperature and heating ramp, growth time, and substrate type. These variables were considered fundamental for reproducible growth, as illustrated in **Figure 4**. Secondary parameters, assigned lower weights, included boat dimensions, tube dimensions, cooling rate, and substrate size, activation, and positioning, which are mainly associated with fine control of growth conditions.

**Figure 5D** shows the methodological evaluation workflow and the results are shown in **Table S10**. After applying the exclusionary criteria, 475 articles remained for methodological evaluation. 115 articles were approved during the methodological evaluation, corresponding to 24% of the studies that passed the abstract screening and only 7% of the initial dataset (**Figure 5E**).

The analysis revealed that most studies prioritized the description of chemical and kinetic parameters. The most frequently reported variables were substrate type (82%), synthesis temperature (80%), carrier gas composition (74%), growth time (68%), and gas flow rate (68%), as shown in **Figure 5F**. Notably, only approximately 50% of the studies reported precursor masses, despite these being among the most fundamental parameters for reproducing any chemical synthesis. In contrast, parameters associated with fluid dynamics and mass transport were frequently neglected. The least reported experimental details included boat dimensions (2%), distance between precursor boats (24%), tube dimensions (27%), and precursor–substrate distance (16%) (**Figure 5F**), highlighting substantial gaps in the reporting of reactor geometry and transport-related conditions.

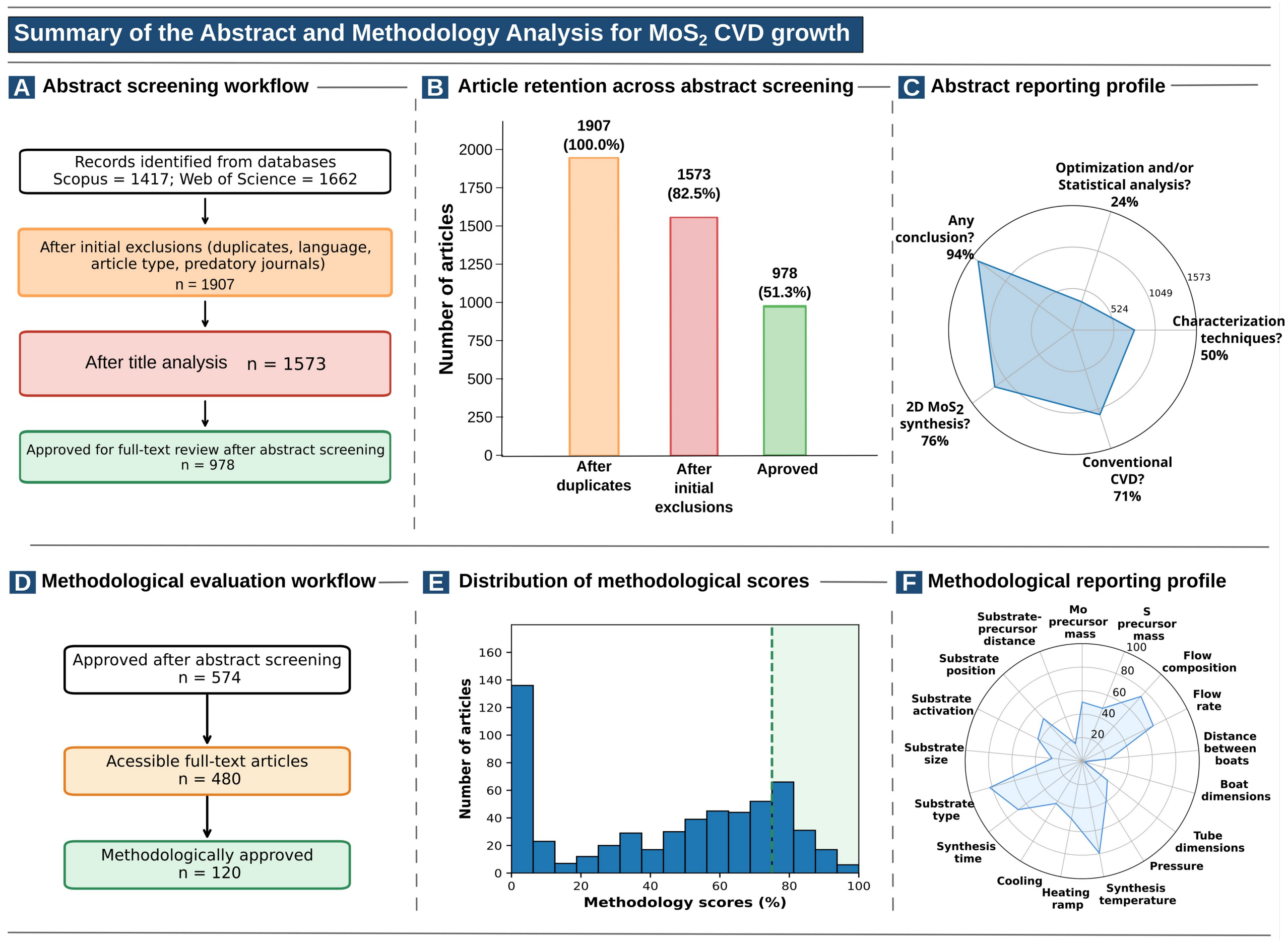

**Figure 5: Summary of the abstract screening and methodological evaluation of publications related to the CVD growth of $MoS_2$.** (A) Workflow of the abstract screening process, showing database retrieval, preliminary exclusions, title analysis, and selection of studies for full-text review. (B) Article retention across the screening stages, illustrating the progressive reduction in the number of studies after applying inclusion and exclusion criteria. (C) Abstract reporting profile based on the weighted screening checklist, highlighting the frequency with which key elements—such as explicit mention of $MoS_2$ synthesis, CVD growth, characterization techniques, statistical analysis, and conclusions—were reported in the abstracts. (D) Workflow of the methodological evaluation, from the articles approved after abstract screening to the final set considered methodologically adequate for meta-analysis. (E) Distribution of methodological scores obtained during the full-text evaluation, with the dashed line indicating the minimum threshold required for approval. (F) Methodological reporting profile summarizing the frequency of reporting of critical experimental parameters related to $MoS_2$ CVD growth, including precursor masses, gas flow conditions, synthesis temperature, reactor geometry, and substrate positioning. The results reveal substantial variability and frequent omissions in the reporting of key parameters affecting synthesis reproducibility.

Although the same general synthesis behavior was described across the approved studies, these articles provided substantially more detailed methodological information. Among the approved works, more than 94% reported key synthesis parameters, including precursor masses, carrier gas composition and flow rate, synthesis temperature, heating rate, growth time, and substrate type and position (**Figure S5**). Parameters associated with reactor fluid dynamics—such as precursor boat distance, tube dimensions, synthesis pressure, and substrate size—were reported in approximately 70% of the studies. Nevertheless, even among articles considered methodologically adequate, important gaps in experimental reporting remained. Parameters directly related to fluid dynamics within the tubular furnace were still frequently neglected, particularly precursor–substrate distance (39%) and boat dimensions (4%), the latter being the least reported parameter overall (**Figure S5**).

These omissions highlight persistent limitations in the reporting of reactor geometry and mass transport conditions, despite their well-established influence on nucleation and crystal growth during CVD synthesis. In CVD growth, even subtle variations in the tubular furnace configuration can significantly alter fluid dynamics, thermal gradients, and precursor transport, which are key factors governing the growth of 2D materials. Under typical flow conditions, parameters such as tube dimensions and precursor boat geometry directly influence gas velocity and precursor distribution, while the relative positioning of the boats modifies flow trajectories and local concentration gradients.[128,129] Consequently, the omission of these parameters compromises the reproducibility of growth kinetics, directly affecting nucleation density, crystal lateral size, morphology, the number of layers, and the defect density of the crystals formed.[129]

The data presented here raise an important warning regarding how methodological sections are reported in the literature, demonstrating that the incomplete description of key experimental parameters limits not only synthesis reproducibility, but also the mechanistic understanding of the growth process itself. Additionally, the lack of detailed reporting for critical parameters may introduce significant bias into meta-analyses, compromising the reliability of cross-study comparisons and statistical interpretations.

## CONCLUSION

This work demonstrates that reproducibility in materials synthesis and characterization remains a major unresolved challenge in materials science, even in systems considered “standard” or widely established. By adapting systematic review methodologies traditionally employed in biomedical sciences to the field of nanomaterials synthesis, we developed a structured framework capable of quantitatively evaluating methodological transparency, reproducibility, and reporting quality across large bodies of literature. Our analysis of Turkevich

AuNP synthesis and CVD-grown $MoS_2$ revealed that, despite decades of accumulated knowledge and thousands of published studies, critical experimental parameters remain inconsistently reported, severely limiting interlaboratory reproducibility and reliable comparison between studies.

The results reinforce that reproducibility problems in materials science are not solely associated with experimental complexity, but are deeply connected to incomplete or selective methodological descriptions and the presence of tacit knowledge within research groups. As discussed by Collins,[130] many experimental procedures depend on implicit practical knowledge that is rarely fully captured in publications, creating a gap between reported protocols and successful experimental replication. Our findings strongly support this perspective. In both case studies, important synthesis variables, such as pH, precursor addition procedures, reactor geometry, gas flow configurations, precursor positioning, and statistical replications were frequently omitted, despite their known influence on nucleation, growth kinetics, morphology, and final material properties.

The present study also aligns with the broader discussion surrounding the scientific reproducibility crisis. Large-scale surveys indicate that most researchers have failed to reproduce published experiments, while selective reporting, insufficient methodological detail, publication pressure, and lack of standardization are consistently identified as major contributing factors. In materials science specifically, recent discussions have emphasized that reproducibility gaps slow technological translation and undermine confidence in emerging materials systems. Our results provide quantitative evidence that these concerns are also highly relevant to nanomaterial synthesis, where subtle experimental variations can profoundly alter material structure and performance.

Importantly, this work highlights that reproducibility should not be interpreted merely as the exact repetition of an experimental procedure, but rather as a broader framework involving transparency, traceability, robustness, and independent validation. The analysis revealed that even studies classified as methodologically acceptable often lacked sufficient detail to enable full replication or meaningful meta-analysis. This limitation becomes even more critical in complex synthesis systems such as CVD growth, where local furnace geometry, thermal gradients, precursor transport, and reactor-specific conditions can strongly affect the resulting materials.

Beyond identifying reproducibility gaps, this work proposes a practical pathway toward improving reliability in materials synthesis research. The combination of systematic review methodologies with computational screening tools demonstrated that large-scale literature evaluation can be performed efficiently and reproducibly. Furthermore, our results reinforce the need for community-wide adoption of standardized reporting practices, including detailed experimental protocols, disclosure of negative or unsuccessful results, statistical treatment of reproducibility, independent validation studies, and structured supplementary datasets. These recommendations are consistent with recent calls in the literature advocating for greater transparency, standardized benchmarking, and reproducibility-oriented publication practices in materials science.[131–133]

Ultimately, this study demonstrates that systematic reviews can serve not only as tools for summarizing scientific literature, but also as powerful instruments for diagnosing structural weaknesses in experimental research practices. The framework proposed here is broadly applicable to other nanomaterials synthesis routes and advanced materials systems, providing a scalable strategy for identifying methodological bottlenecks, improving experimental transparency, and strengthening the reliability and long-term impact of materials science research.

## ASSOCIATED CONTENT

The Supporting Information includes the complete search strategies used for the Scopus and Web of Science databases, the adapted STROBE-based checklists employed during the abstract and methodological screening stages, the weighting criteria assigned to each evaluation parameter, and the validation procedures used for the screening algorithm. It also contains the full methodological evaluation tables, reporting profiles, supplementary figures, and detailed discussions regarding the development and optimization of the classification workflow for both the AuNP and $MoS_2$ case studies. Additionally, the complete screening and classification algorithm, together with all tables containing the results from each stage of the systematic review and meta-analysis workflow, are publicly available through the Open Science Framework (OSF) repository (https://doi.org/10.17605/OSF.IO/YSPHK). This repository ensures full transparency, reproducibility, and accessibility of the data analysis procedures used in this study.

## AUTHOR INFORMATION

**ORCID**

Julia S. Correa: https://orcid.org/0000-0003-1522-3795

Alex S. Lima: https://orcid.org/0000-0003-1727-2646

Daniel Grasseschi: https://orcid.org/0000-0001-6066-0869

**Corresponding Author**

*Daniel Grasseschi. Emial: dgrasseschi@iq.ufrj.br

**Present Addresses**

Department of Inorganic Chemistry, Chemistry Institute – Federal University of Rio de Janeiro Av. Pedro Calmon, 550 – CidadeUniversitária, 21941-901, Rio de Janeiro, Brazil.

**Author Contributions**

The manuscript was written through contributions from all authors, and all authors approved the final version of the manuscript. ‡These authors contributed equally to this work. JSC and DG conducted the abstract screening of the AuNP articles. JSC performed the methodological screening and meta-analysis of the AuNP dataset. LVS and NG conducted the systematic review, methodological evaluation, and meta-analysis of the $MoS_2$ CVD growth studies. CRJ developed, optimized, and applied the screening and classification algorithm. The entire study was supervised by DG.

**Funding Sources**

This work was supported by the the Brazilian Nanocarbon Institute of Science and Technology (INCT/Nanocarbono), Conselho Nacional de Desenvolvimento Científico e Tecnológico (CNPq), and Coordenação de Aperfeiçoamento de Pessoal de Nível Superior (CAPES). DG acknowledges financial support from Rio de Janeiro State Foundation (FAPERJ – grant nos. E-26/210.296/2022, E-26/211.464/2021, and E-26/201.254/2022), Serrapilheira Institute (grant number Serra – R-2012-37959), and Financiadora de Estudos e Projetos – Finep grant number 01.22.0208.00.

**ACKNOWLEDGMENT**

The authors acknowledge the support of the Brazilian Reproducibility Network for promoting discussions, initiatives, and collaborative efforts aimed at improving transparency, methodological rigor, and reproducibility in scientific research. The discussions and resources provided by the network were valuable for the development of the systematic review framework and for fostering critical reflections on reproducibility challenges in materials science.

This work was supported by the the Brazilian Nanocarbon Institute of Science and Technology (INCT/Nanocarbono), Conselho Nacional de Desenvolvimento Científico e Tecnológico (CNPq), and Coordenação de Aperfeiçoamento de Pessoal de Nível Superior (CAPES). DG acknowledges financial support from Rio de Janeiro State Foundation (FAPERJ – grant nos. E-26/210.296/2022, E-26/211.464/2021, and E-26/201.254/2022), Serrapilheira Institute (grant number Serra – R-2012-37959), and Financiadora de Estudos e Projetos – Finep grant number 01.22.0208.00.

The authors used OpenAI ChatGPT-5.5 to assist with Python code debugging, modify visual data, and refining language, spell check, and grammar correction. All content has been reviewed and edited by the authors, who take full responsibility for the accuracy and originality of the work.

**REFERENCES**

(1) Stoddart, C. Is There a Reproducibility Crisis in Science? *Nature* **2016**. https://doi.org/10.1038/d41586-019-00067-3.

(2) The Brazilian Reproducibility Initiative; Amaral, O. B.; Carneiro, C. F. D.; Neves, K.; Sampaio, A. P. W.; Gomes, B. V.; Abreu, M. B. D.; Tan, P. B.; Mota, G. P. S.; Goulart, R. N.; Fernandes, N. R. D. S.; Linhares, J. H.; Ibelli, A. M. G.; Sebollela, A.; Andricopulo, A. D.; Carvalho, A. A. V. D.; Silva, A. P. E.; Souza, A. S. O.; Lima, A. C. C. D.; Sousa, A. M. D.; Birbrair, A.; Borbely, A. U.; Machado, A. M.; Costa, A. D. C.; Vasconcelos, A. P. D.; Silva, A. H. B. D. L.; Souza, A. D.; Herrmann, A. P.; Bastos, A. P. A.; Nunes, A. C. C.; Castrucci, A. M. D. L.; Paula, A. B. R.; Waltrick, A. P. F.; Macedo, A. G. F. D.; Mecawi, A. S.; Rinaldi, A. W.; Canedo, A. D.; Vidigal, A. P. P.; Marcal, A. P.; Monteiro, A. M.; Rocha, B. G. S.; Pachane, B. C.; Andrade, B. D. S.; Casali, B. C.; Popik, B.; Souza, C. P. D.; Santos, C. D. S. D.; Goncalves, C. M.; Nascimento, C. R. B. D.; Verissimo, C. P.; Girardi, C. E. N.; Penido, C.; Batista, C.; Saibro-Girardi, C.; Pedrosa, C. D. S. G.; Panis, C.; Picoli, C. D. C.; Furini, C. R. G.; Caetano, D. S. L.; Gelain, D. P.; Silva, D. C.; Aguiar, D. C. D.; Goncalves, D. A.; Silva, D. S. D.; Araujo, D. A. M. D.; Junior, D. A. C.; Campos, D. B. P. D.; Nachtigall, E. G.; Tanabe, E. L. L.; Kioshima, E. S.; Seiva, F. R. F.; Mendes, F. D. A.; Silva, F. J. M. D.; Moreira, F. D. A.; Vanz, F.; Leser, F. S.; Ferrao, F. M.; Lotz, F. N.; Silva, F. C. D. S.; Bloise, F. F.; Lima, F. R. S.; Lara, F. A.; Volpato, F. A.; Stefanello, F. M.; Vitorino, F. N. D. L.; Fonseca, F. N.; Furtado, G. V.; Neves, G. A.; Zanetti, G.; Cancelliero, G. S.; Pisani, G. F. D.; Casagrande, G. C. A.; Villas-Boas, G. R.; Cruz, H. R. O. A. D.; Borges, H. L.; Araujo, H. S. S. D.; Povoa, H. C. C.; Brunel, H. D. S. S.; Bayer, H.; Baptista, I. L.; Werle, I.; Divino, I. A.; Jardim, I. A. B. A.; Zanoveli, J. M.; Peixoto, J. D. O.; Rinaldi, J. D. C.; Ferreira, J. B.; Franco, J. L.; Rugani, J. M. N.; Cruz, J. V. R.; Myskiw, J. D. C.; Quillfeldt, J. A.; Costa, J. M. J. P. D.; Barros, J. H. O. D.; Barros, J. M.; Cunha, J. P. C. D.; Ayub, J. G. M.; Freitas, J. D. R. D.; Roson, J. N.; Borbely, K. S. C.; Asensi, K. D.; Queiroz, K. B. D.; Rodrigues, K. D. S.; Vieira, K. D. C.; Guerra, K. T. K.; Pires, K. S. N.; Trajano-Silva, L. A. M.; Bertoglio, L. J.; Assis, L. V. M. D.; Lery, L. M. S.; Vasconcelos, L. M.; Silva, L. D. O. M. D.; Fernandes-Siqueira, L. D. O.; Alvares, L. D. O.; Falcione, L. H. D. M.; Silva, L. D. S. D.; Vieira, L. G.; Madeira, L. F.; Nunes, L. E. D.; Chuffa, L. G. D. A.; Behrens, L. M. P.; Esteca, M. V.; Felipe, M. S. S.; Moraes, M. N.; Saraiva-Pereira, M. L.; Silveira, M. S. D.; Duarte, M. S. V.; Porcionatto, M. A.; Salvi, M.; Clavero, M. A.; Ferreira, M. L.; Gallas-Lopes, M.; Moura, M. A. D.; Isoldi, M. C.; Mundim, M. T. V. V.; Andrades, M.; Donato, M. F.; Barboza, M. P.; Fagundes, M. M. D. A.; Silva, M. A.; Ledur, M. C.; Silva, M. D. D.; Baqui, M. M. A.; Souza, M. P. D.; Povoa, N. I. L. P.; Bellini, N. K.; Pedra, N. S.; Somensi, N.; Soares, N. P.; Nunes, N. M. P.; Silva, N. F.; Carvalho, P. M. G. D.; Cappelletti, P. A.; Garcez, P. P.; Martins, P. R.; Brito, P. K. M. O.; Lauar, P. R. M.; Nicolicht-Amorim, P.; Chitolina, R.; Barbosa, R. A. Q.; Cordova, R. A. G.; Goldenberg, R. C. D. S.; Tavares, R. G.; Saito, R. D. F.; Krogh, R.; Andreatini, R.; Almeida, R. F.; Menezes, R. C. A. D.; Rorato, R.; Chammas, R.; Andrade, R. V. D.; Spanevello, R. M.; Costa, R. A.; Monte-Neto, R. L. D.; Reis, S. A. D.; Kraus, S. I.; Silva, S. C.; Michelan-Duarte, S.; Rehen, S. K.; Noronha, S. I. S. R. D.; Ortiga-Carvalho, T. M.; Prieto, T. D. S.; Fagundes, T. R.; Nakamura, T. K. E.; Costa, T. E. M. M.; Junior, T. C. T.; Luz, T. B. D.; Silva, T. A. D.; Santos, T. G. D.; Beijamini, V.; Monteiro, V. R. D. S.; Wippel, V. A.; Valenca, V. M. O.; Altei, W. F. Estimating the Replicability of Brazilian Biomedical Science. Cold Spring Harbor Laboratory April 3, 2025. https://doi.org/10.1101/2025.04.02.645026.

(3) Persaud, D.; Ward, L.; Hattrick-Simpers, J. Reproducibility in Materials Informatics: Lessons from 'A General-Purpose Machine Learning Framework for Predicting Properties of Inorganic Materials.' *Digital Discovery* **2024**, *3* (2), 281–286. https://doi.org/10.1039/d3dd00199g.

(4) Bøggild, P.; Booth, T. J.; Lassaline, N.; Jessen, B. S.; Shivayogimath, A.; Hofmann, S.; Daasbjerg, K.; Smith, A.; Nørgaard, K.; Zurutuza, A.; Asselberghs, I.; Barkan, T.; Taboryski, R.; Pollard, A. J. Closing the Reproducibility Gap: 2D Materials Research. arXiv 2024. https://doi.org/10.48550/ARXIV.2409.18994.

(5) Bøggild, P. Research on Scalable Graphene Faces a Reproducibility Gap. *Nat Commun* **2023**, *14* (1). https://doi.org/10.1038/s41467-023-36891-5.

(6) PRISMA-P Group; Moher, D.; Shamseer, L.; Clarke, M.; Ghersi, D.; Liberati, A.; Petticrew, M.; Shekelle, P.; Stewart, L. A. Preferred Reporting Items for Systematic Review and Meta-Analysis Protocols (PRISMA-P) 2015 Statement. *Syst Rev* **2015**, *4* (1), 1. https://doi.org/10.1186/2046-4053-4-1.

(7) Siddaway, A. P.; Wood, A. M.; Hedges, L. V. How to Do a Systematic Review: A Best Practice Guide for Conducting and Reporting Narrative Reviews, Meta-Analyses, and Meta-Syntheses. *Annu. Rev. Psychol.* **2019**, *70* (1), 747–770. https://doi.org/10.1146/annurev-psych-010418-102803.
(8) Ahn, E.; Kang, H. Introduction to Systematic Review and Meta-Analysis. *Korean J Anesthesiol* **2018**, *71* (2), 103–112. https://doi.org/10.4097/kjae.2018.71.2.103.
(9) Phillips, V.; Barker, E. Systematic Reviews: Structure, Form and Content. *Journal of Perioperative Practice* **2021**, *31* (9), 349–353. https://doi.org/10.1177/1750458921994693.
(10) Owens, J. K. Systematic Reviews: Brief Overview of Methods, Limitations, and Resources. *Nurse Author & Editor* **2021**, *31* (3–4), 69–72. https://doi.org/10.1111/nae2.28.
(11) Wu, W.; Duan, J.; Reed, W. R.; Tipton, E. What Can We Learn from 1,000 Meta-Analyses across 10 Different Disciplines? *Res. synth. methods* **2026**, *17* (1), 123–156. https://doi.org/10.1017/rsm.2025.10035.
(12) Nakagawa, S.; Yang, Y.; Macartney, E. L.; Spake, R.; Lagisz, M. Quantitative Evidence Synthesis: A Practical Guide on Meta-Analysis, Meta-Regression, and Publication Bias Tests for Environmental Sciences. *Environ Evid* **2023**, *12* (1), 8. https://doi.org/10.1186/s13750-023-00301-6.
(13) Ahmad, Z.; Ammar, M.; Sellami, A.; Al-Thani, N. J. Effective Pedagogical Approaches Used in High School Chemistry Education: A Systematic Review and Meta-Analysis. *J. Chem. Educ.* **2023**, *100* (5), 1796–1810. https://doi.org/10.1021/acs.jchemed.2c00739.
(14) Higgins, J.; Thomas, J.; Chandler, J.; Li, T.; Page, M. J. *Cochrane Handbook for Systematic Reviews of Interventions*, Version 6.5.; 2024.
(15) National Institute for Health and Care Research. *PROSPERO International prospective register of systematic reviews*. https://www.crd.york.ac.uk/prospero/.
(16) Cooke, A.; Smith, D.; Booth, A. Beyond PICO: The SPIDER Tool for Qualitative Evidence Synthesis. *Qual Health Res* **2012**, *22* (10), 1435–1443. https://doi.org/10.1177/1049732312452938.
(17) Grasseschi, D.; De O. Pereira, M. L.; Shinohara, J. S.; Toma, H. E. Facile Synthesis of Labile Gold Nanodiscs by the Turkevich Method. *J Nanopart Res* **2018**, *20* (2), 35. https://doi.org/10.1007/s11051-018-4149-y.
(18) Shi, L.; Buhler, E.; Boué, F.; Carn, F. How Does the Size of Gold Nanoparticles Depend on Citrate to Gold Ratio in Turkevich Synthesis? Final Answer to a Debated Question. *Journal of Colloid and Interface Science* **2017**, *492*, 191–198. https://doi.org/10.1016/j.jcis.2016.10.065.
(19) Wuithschick, M.; Birnbaum, A.; Witte, S.; Sztucki, M.; Vainio, U.; Pinna, N.; Rademann, K.; Emmerling, F.; Kraehnert, R.; Polte, J. Turkevich in New Robes: Key Questions Answered for the Most Common Gold Nanoparticle Synthesis. *ACS Nano* **2015**, *9* (7), 7052–7071. https://doi.org/10.1021/acsnano.5b01579.
(20) Grasseschi, D.; Ando, R. A.; Toma, H. E.; Zamarion, V. M. Unraveling the Nature of Turkevich Gold Nanoparticles: The Unexpected Role of the Dicarboxyketone Species. *RSC Adv.* **2015**, *5* (8), 5716–5724. https://doi.org/10.1039/C4RA12161A.
(21) Schulz, F.; Homolka, T.; Bastús, N. G.; Puntes, V.; Weller, H.; Vossmeyer, T. Little Adjustments Significantly Improve the Turkevich Synthesis of Gold Nanoparticles. *Langmuir* **2014**, *30* (35), 10779–10784. https://doi.org/10.1021/la503209b.
(22) Ojea-Jiménez, I.; Bastús, N. G.; Puntes, V. Influence of the Sequence of the Reagents Addition in the Citrate-Mediated Synthesis of Gold Nanoparticles. *J. Phys. Chem. C* **2011**, *115* (32), 15752–15757. https://doi.org/10.1021/jp2017242.
(23) Bastús, N. G.; Comenge, J.; Puntes, V. Kinetically Controlled Seeded Growth Synthesis of Citrate-Stabilized Gold Nanoparticles of up to 200 Nm: Size Focusing versus Ostwald Ripening. *Langmuir* **2011**, *27* (17), 11098–11105. https://doi.org/10.1021/la201938u.
(24) Xia, H.; Bai, S.; Hartmann, J.; Wang, D. Synthesis of Monodisperse Quasi-Spherical Gold Nanoparticles in Water via Silver(I)-Assisted Citrate Reduction. *Langmuir* **2010**, *26* (5), 3585–3589. https://doi.org/10.1021/la902987w.
(25) Polte, J.; Ahner, T. T.; Delissen, F.; Sokolov, S.; Emmerling, F.; Thünemann, A. F.; Kraehnert, R. Mechanism of Gold Nanoparticle Formation in the Classical Citrate Synthesis Method Derived from Coupled In Situ XANES and SAXS Evaluation. *J. Am. Chem. Soc.* **2010**, *132*

(4), 1296–1301. https://doi.org/10.1021/ja906506j.
(26) Liu, Z.; Zu, Y.; Guo, S. Synthesis of Micron-Scale Gold Nanochains by a Modified Citrate Reduction Method. *Applied Surface Science* **2009**, *255* (11), 5827–5830. https://doi.org/10.1016/j.apsusc.2009.01.012.
(27) Patungwasa, W.; Hodak, J. H. pH Tunable Morphology of the Gold Nanoparticles Produced by Citrate Reduction. *Materials Chemistry and Physics* **2008**, *108* (1), 45–54. https://doi.org/10.1016/j.matchemphys.2007.09.001.
(28) Pong, B.-K.; Elim, H. I.; Chong, J.-X.; Ji, W.; Trout, B. L.; Lee, J.-Y. New Insights on the Nanoparticle Growth Mechanism in the Citrate Reduction of Gold(III) Salt: Formation of the Au Nanowire Intermediate and Its Nonlinear Optical Properties. *J. Phys. Chem. C* **2007**, *111* (17), 6281–6287. https://doi.org/10.1021/jp068666o.
(29) Ji, X.; Song, X.; Li, J.; Bai, Y.; Yang, W.; Peng, X. Size Control of Gold Nanocrystals in Citrate Reduction: The Third Role of Citrate. *J. Am. Chem. Soc.* **2007**, *129* (45), 13939–13948. https://doi.org/10.1021/ja074447k.
(30) Kimling, J.; Maier, M.; Okenve, B.; Kotaidis, V.; Ballot, H.; Plech, A. Turkevich Method for Gold Nanoparticle Synthesis Revisited. *J. Phys. Chem. B* **2006**, *110* (32), 15700–15707. https://doi.org/10.1021/jp061667w.
(31) Turkevich, J.; Stevenson, P. C.; Hillier, J. A Study of the Nucleation and Growth Processes in the Synthesis of Colloidal Gold. *Discuss. Faraday Soc.* **1951**, *11*, 55. https://doi.org/10.1039/df9511100055.
(32) Schulz, F.; Homolka, T.; Bastús, N. G.; Puntes, V.; Weller, H.; Vossmeyer, T. Little Adjustments Significantly Improve the Turkevich Synthesis of Gold Nanoparticles. *Langmuir* **2014**, *30* (35), 10779–10784. https://doi.org/10.1021/la503209b.
(33) Personick, M. L.; Mirkin, C. A. Making Sense of the Mayhem behind Shape Control in the Synthesis of Gold Nanoparticles. *J. Am. Chem. Soc.* **2013**, *135* (49), 18238–18247. https://doi.org/10.1021/ja408645b.
(34) Kimling, J.; Maier, M.; Okenve, B.; Kotaidis, V.; Ballot, H.; Plech, A. Turkevich Method for Gold Nanoparticle Synthesis Revisited. *J. Phys. Chem. B* **2006**, *110* (32), 15700–15707. https://doi.org/10.1021/jp061667w.
(35) Polte, J.; Ahner, T. T.; Delissen, F.; Sokolov, S.; Emmerling, F.; Thünemann, A. F.; Kraehnert, R. Mechanism of Gold Nanoparticle Formation in the Classical Citrate Synthesis Method Derived from Coupled In Situ XANES and SAXS Evaluation. *J. Am. Chem. Soc.* **2010**, *132* (4), 1296–1301. https://doi.org/10.1021/ja906506j.
(36) Lee, Y.-H.; Zhang, X.-Q.; Zhang, W.; Chang, M.-T.; Lin, C.-T.; Chang, K.-D.; Yu, Y.-C.; Wang, J. T.-W.; Chang, C.-S.; Li, L.-J.; Lin, T.-W. Synthesis of Large-Area MoS 2 Atomic Layers with Chemical Vapor Deposition. *Advanced Materials* **2012**, *24* (17), 2320–2325. https://doi.org/10.1002/adma.201104798.
(37) Balendhran, S.; Ou, J. Z.; Bhaskaran, M.; Sriram, S.; Ippolito, S.; Vasic, Z.; Kats, E.; Bhargava, S.; Zhuiykov, S.; Kalantar-zadeh, K. Atomically Thin Layers of MoS 2 via a Two Step Thermal Evaporation–Exfoliation Method. *Nanoscale* **2012**, *4* (2), 461–466. https://doi.org/10.1039/C1NR10803D.
(38) Ling, X.; Lee, Y.-H.; Lin, Y.; Fang, W.; Yu, L.; Dresselhaus, M. S.; Kong, J. Role of the Seeding Promoter in MoS 2 Growth by Chemical Vapor Deposition. *Nano Letters* **2014**, *14* (2), 464–472. https://doi.org/10.1021/nl4033704.
(39) Jeon, J.; Jang, S. K.; Jeon, S. M.; Yoo, G.; Jang, Y. H.; Park, J.-H.; Lee, S. Layer-Controlled CVD Growth of Large-Area Two-Dimensional $MoS_2$ Films. *Nanoscale* **2015**, *7* (5), 1688–1695. https://doi.org/10.1039/C4NR04532G.
(40) Kang, K.; Xie, S.; Huang, L.; Han, Y.; Huang, P. Y.; Mak, K. F.; Kim, C.-J.; Muller, D.; Park, J. High-Mobility Three-Atom-Thick Semiconducting Films with Wafer-Scale Homogeneity. *Nature* **2015**, *520* (7549), 656–660. https://doi.org/10.1038/nature14417.
(41) Wang, W.; Zeng, X.; Wu, S.; Zeng, Y.; Hu, Y.; Ding, J.; Xu, S. Effect of Mo Concentration on Shape and Size of Monolayer $MoS_2$ Crystals by Chemical Vapor Deposition. *J. Phys. D: Appl. Phys.* **2017**, *50* (39), 395501. https://doi.org/10.1088/1361-6463/aa81ae.
(42) Zhu, D.; Shu, H.; Jiang, F.; Lv, D.; Asokan, V.; Omar, O.; Yuan, J.; Zhang, Z.; Jin, C. Capture the Growth Kinetics of CVD Growth of Two-Dimensional MoS2. *npj 2D Mater Appl* **2017**, *1* (1), 8. https://doi.org/10.1038/s41699-017-0010-x.

(43) Han, T.; Liu, H.; Wang, S.; Li, W.; Chen, S.; Yang, X.; Cai, M. Research on the Factors Affecting the Growth of Large-Size Monolayer MoS2 by APCVD. *Materials* **2018**, *11* (12), 2562. https://doi.org/10.3390/ma11122562.
(44) Li, X.; Kahn, E.; Chen, G.; Sang, X.; Lei, J.; Passarello, D.; Oyedele, A. D.; Zakhidov, D.; Chen, K.-W.; Chen, Y.-X.; Hsieh, S.-H.; Fujisawa, K.; Unocic, R. R.; Xiao, K.; Salleo, A.; Toney, M. F.; Chen, C.-H.; Kaxiras, E.; Terrones, M.; Yakobson, B. I.; Harutyunyan, A. R. Surfactant-Mediated Growth and Patterning of Atomically Thin Transition Metal Dichalcogenides. *ACS Nano* **2020**, *14* (6), 6570–6581. https://doi.org/10.1021/acsnano.0c00132.
(45) Seravalli, L.; Bosi, M. A Review on Chemical Vapour Deposition of Two-Dimensional MoS2 Flakes. *Materials* **2021**, *14* (24), 7590. https://doi.org/10.3390/ma14247590.
(46) Deng, Y.; Ge, L.; Tang, J. Synthesis, Functionalization and Assembly of Monodisperse Gold Nanoparticles with Potassium Sodium Tartrate as Reducing Agent. *ChemistrySelect* **2024**, *9* (8), e202304906. https://doi.org/10.1002/slct.202304906.
(47) Kettemann, F.; Birnbaum, A.; Witte, S.; Wuithschick, M.; Pinna, N.; Kraehnert, R.; Rademann, K.; Polte, J. Missing Piece of the Mechanism of the Turkevich Method: The Critical Role of Citrate Protonation. *Chem. Mater.* **2016**, *28* (11), 4072–4081. https://doi.org/10.1021/acs.chemmater.6b01796.
(48) Witzigmann, D.; Sieber, S.; Porta, F.; Grossen, P.; Bieri, A.; Strelnikova, N.; Pfohl, T.; Prescianotto-Baschong, C.; Huwyler, J. Formation of Lipid and Polymer Based Gold Nanohybrids Using a Nanoreactor Approach. *RSC Adv.* **2015**, *5* (91), 74320–74328. https://doi.org/10.1039/C5RA13967H.
(49) Bayazit, M. K.; Yue, J.; Cao, E.; Gavriilidis, A.; Tang, J. Controllable Synthesis of Gold Nanoparticles in Aqueous Solution by Microwave Assisted Flow Chemistry. *ACS Sustainable Chem. Eng.* **2016**, *4* (12), 6435–6442. https://doi.org/10.1021/acssuschemeng.6b01149.
(50) Oliveira, J. P.; Prado, A. R.; Keijok, W. J.; Ribeiro, M. R. N.; Pontes, M. J.; Nogueira, B. V.; Guimarães, M. C. C. A Helpful Method for Controlled Synthesis of Monodisperse Gold Nanoparticles through Response Surface Modeling. *Arabian Journal of Chemistry* **2020**, *13* (1), 216–226. https://doi.org/10.1016/j.arabjc.2017.04.003.
(51) Bari, A. H.; Shukla, N.; Gavriilidis, A.; Kulkarni, A. A. Transient Response to Perturbations in Flow Synthesis of Citrate Capped Gold Nanoparticles. *Chemical Engineering Journal* **2023**, *470*, 143890. https://doi.org/10.1016/j.cej.2023.143890.
(52) Manta, P.; Chandra Singh, S.; Deep, A.; Kapoor, D. N. Temperature-regulated Gold Nanoparticle Sensors for Immune Chromatographic Rapid Test Kits with Reproducible Sensitivity: A Study. *IET Nanobiotechnology* **2021**, *15* (3), 338–346. https://doi.org/10.1049/nbt2.12024.
(53) Cvak, B.; Pum, D.; Molinelli, A.; Krska, R. Synthesis and Characterization of Colloidal Gold Particles as Labels for Antibodies as Used in Lateral Flow Devices. *Analyst* **2012**, *137* (8), 1882. https://doi.org/10.1039/c2an16108g.
(54) Stein, R.; Friedrich, B.; Mühlberger, M.; Cebulla, N.; Schreiber, E.; Tietze, R.; Cicha, I.; Alexiou, C.; Dutz, S.; Boccaccini, A. R.; Unterweger, H. Synthesis and Characterization of Citrate-Stabilized Gold-Coated Superparamagnetic Iron Oxide Nanoparticles for Biomedical Applications. *Molecules* **2020**, *25* (19), 4425. https://doi.org/10.3390/molecules25194425.
(55) Cornwell, M.; Damilos, S.; Parkin, I. P.; Gavriilidis, A. Seeded-Growth Synthesis of 20–60 Nm Monodisperse Citrate-Capped Gold Nanoparticles in a Millifluidic Reactor. *J Flow Chem* **2024**, *14* (4), 655–666. https://doi.org/10.1007/s41981-024-00334-z.
(56) Larm, N. E.; Essner, J. B.; Pokpas, K.; Canon, J. A.; Jahed, N.; Iwuoha, E. I.; Baker, G. A. Room-Temperature Turkevich Method: Formation of Gold Nanoparticles at the Speed of Mixing Using Cyclic Oxocarbon Reducing Agents. *J. Phys. Chem. C* **2018**, *122* (9), 5105–5118. https://doi.org/10.1021/acs.jpcc.7b10536.
(57) Ali, M. M.; A. Rajab, N.; A. Abdulrasool, A. Preparation, Characterization and Optimization of Etoposide-Loaded Gold Nanoparticles Based on Chemical Reduction Method. *IJPS* **2020**, *29* (2), 107–121. https://doi.org/10.31351/vol29iss2pp107-121.
(58) Díaz-García, V.; Haensgen, A.; Inostroza, L.; Contreras-Trigo, B.; Oyarzun, P. Novel Microsynthesis of High-Yield Gold Nanoparticles to Accelerate Research in Biosensing and Other Bioapplications. *Biosensors* **2023**, *13* (12), 992. https://doi.org/10.3390/bios13120992.
(59) Panariello, L.; Damilos, S.; Du Toit, H.; Wu, G.; Radhakrishnan, A. N. P.; Parkin, I. P.;

Gavriilidis, A. Highly Reproducible, High-Yield Flow Synthesis of Gold Nanoparticles Based on a Rational Reactor Design Exploiting the Reduction of Passivated Au( III ). *React. Chem. Eng.* **2020**, *5* (4), 663–676. https://doi.org/10.1039/C9RE00469F.
(60) Rawat, P.; Imam, S. S.; Gupta, S. Formulation of Cabotegravir Loaded Gold Nanoparticles: Optimization, Characterization to In-Vitro Cytotoxicity Study. *J Clust Sci* **2023**, *34* (2), 893–905. https://doi.org/10.1007/s10876-022-02261-2.
(61) Grasseschi, D.; De O. Pereira, M. L.; Shinohara, J. S.; Toma, H. E. Facile Synthesis of Labile Gold Nanodiscs by the Turkevich Method. *J Nanopart Res* **2018**, *20* (2), 35. https://doi.org/10.1007/s11051-018-4149-y.
(62) Oluwatosin Kudirat, S.; Tawakalitu, A.; A. Saka, A.; O. Kamaldeen, A.; Mercy T, B.; Jimoh Oladejo, T. Entrapped Chemically Synthesized Gold Nanoparticles Combined with Polyethylene Glycol and Chloroquine Diphosphate as an Improved Antimalarial Drug. *Nanomed J* **2019**, *6* (2). https://doi.org/10.22038/nmj.2019.06.0002.
(63) Bahmanyar, Z.; Mohammadi, F.; Gholami, A.; Khoshneviszadeh, M. Effect of Different Physical Factors on the Synthesis of Spherical Gold Nanoparticles towards Cost-effective Biomedical Applications. *IET Nanobiotechnology* **2023**, *17* (1), 1–12. https://doi.org/10.1049/nbt2.12100.
(64) Choi, J. W.; Kim, Y. J.; Lee, J. M.; Choi, J.-H.; Choi, J.-W.; Chung, B. G. Droplet-Based Synthesis of Homogeneous Gold Nanoparticles for Enhancing HRP-Based ELISA Signals. *BioChip J* **2020**, *14* (3), 298–307. https://doi.org/10.1007/s13206-020-4307-z.
(65) Dong, J.; Lau, J.; Svoronos, S. A.; Moudgil, B. M. Continuous Synthesis of Precision Gold Nanoparticles Using a Flow Reactor. *KONA* **2022**, *39* (0), 262–269. https://doi.org/10.14356/kona.2022011.
(66) Heiserer, S.; Eder, P.; Ó Coileáin, C.; Biba, J.; Stimpel-Lindner, T.; Bartlam, C.; Rührmair, U.; Duesberg, G. S. Controllable and Reproducible Growth of Transition Metal Dichalcogenides by Design of Experiments. *Adv Elect Materials* **2023**, *9* (10), 2300281. https://doi.org/10.1002/aelm.202300281.
(67) Zhou, J.; Lin, J.; Huang, X.; Zhou, Y.; Chen, Y.; Xia, J.; Wang, H.; Xie, Y.; Yu, H.; Lei, J.; Wu, D.; Liu, F.; Fu, Q.; Zeng, Q.; Hsu, C.-H.; Yang, C.; Lu, L.; Yu, T.; Shen, Z.; Lin, H.; Yakobson, B. I.; Liu, Q.; Suenaga, K.; Liu, G.; Liu, Z. A Library of Atomically Thin Metal Chalcogenides. *Nature* **2018**, *556* (7701), 355–359. https://doi.org/10.1038/s41586-018-0008-3.
(68) Xu, H.; Zhou, W.; Zheng, X.; Huang, J.; Feng, X.; Ye, L.; Xu, G.; Lin, F. Control of the Nucleation Density of Molybdenum Disulfide in Large-Scale Synthesis Using Chemical Vapor Deposition. *Materials* **2018**, *11* (6), 870. https://doi.org/10.3390/ma11060870.
(69) Xie, Y.; Wang, Z.; Zhan, Y.; Zhang, P.; Wu, R.; Jiang, T.; Wu, S.; Wang, H.; Zhao, Y.; Nan, T.; Ma, X. Controllable Growth of Monolayer $MoS_2$ by Chemical Vapor Deposition via Close $MoO_2$ Precursor for Electrical and Optical Applications. *Nanotechnology* **2017**, *28* (8), 084001. https://doi.org/10.1088/1361-6528/aa5439.
(70) Goto, Y.; Ogino, A. CVD Synthesis of Monolayer $MoS_2$ Using Na Compounds as Additives to Enhance Two-Dimensional Growth. *Jpn. J. Appl. Phys.* **2023**, *62* (7), 075503. https://doi.org/10.35848/1347-4065/ace397.
(71) Ponnusamy, K.; Durairaj, S.; Chandramohan, S. Effect of Growth Temperature on the Morphology Control and Optical Behavior of Monolayer MoS2 on SiO2 Substrate. *J Mater Sci: Mater Electron* **2022**, *33* (12), 9549–9557. https://doi.org/10.1007/s10854-021-07547-1.
(72) Yang, S. Y.; Shim, G. W.; Seo, S.-B.; Choi, S.-Y. Effective Shape-Controlled Growth of Monolayer MoS2 Flakes by Powder-Based Chemical Vapor Deposition. *Nano Res.* **2017**, *10* (1), 255–262. https://doi.org/10.1007/s12274-016-1284-6.
(73) Narayanan, V.; M A, G.; Rahman, A. How to ‘Train’ Your CVD to Grow Large-Area 2D Materials. *Mater. Res. Express* **2019**, *6* (12), 125002. https://doi.org/10.1088/2053-1591/ab5383.
(74) Dumcenco, D.; Ovchinnikov, D.; Marinov, K.; Lazić, P.; Gibertini, M.; Marzari, N.; Sanchez, O. L.; Kung, Y.-C.; Krasnozhon, D.; Chen, M.-W.; Bertolazzi, S.; Gillet, P.; Fontcuberta I Morral, A.; Radenovic, A.; Kis, A. Large-Area Epitaxial Monolayer $MoS_2$. *ACS Nano* **2015**, *9* (4), 4611–4620. https://doi.org/10.1021/acsnano.5b01281.
(75) Singh, A.; Sharma, M.; Singh, R. NaCl-Assisted CVD Growth of Large-Area High-Quality Trilayer $MoS_2$ and the Role of the Concentration Boundary Layer. *Crystal Growth & Design*

**2021**, *21* (9), 4940–4946. https://doi.org/10.1021/acs.cgd.1c00390.
(76) Suleman, M.; Lee, S.; Kim, M.; Nguyen, V. H.; Riaz, M.; Nasir, N.; Kumar, S.; Park, H. M.; Jung, J.; Seo, Y. NaCl-Assisted Temperature-Dependent Controllable Growth of Large-Area $MoS_2$ Crystals Using Confined-Space CVD. *ACS Omega* **2022**, *7* (34), 30074–30086. https://doi.org/10.1021/acsomega.2c03108.
(77) Ling, X.; Lee, Y.-H.; Lin, Y.; Fang, W.; Yu, L.; Dresselhaus, M. S.; Kong, J. Role of the Seeding Promoter in $MoS_2$ Growth by Chemical Vapor Deposition. *Nano Lett.* **2014**, *14* (2), 464–472. https://doi.org/10.1021/nl4033704.
(78) Lu, Y.; Chen, T.; Ryu, G. H.; Huang, H.; Sheng, Y.; Chang, R.-J.; Warner, J. H. Self-Limiting Growth of High-Quality 2D Monolayer $MoS_2$ by Direct Sulfurization Using Precursor-Soluble Substrates for Advanced Field-Effect Transistors and Photodetectors. *ACS Appl. Nano Mater.* **2019**, *2* (1), 369–378. https://doi.org/10.1021/acsanm.8b01955.
(79) Wang, S.; Rong, Y.; Fan, Y.; Pacios, M.; Bhaskaran, H.; He, K.; Warner, J. H. Shape Evolution of Monolayer $MoS_2$ Crystals Grown by Chemical Vapor Deposition. *Chem. Mater.* **2014**, *26* (22), 6371–6379. https://doi.org/10.1021/cm5025662.
(80) Cho, Y. J.; Sim, Y.; Lee, J.-H.; Hoang, N. T.; Seong, M.-J. Size and Shape Control of CVD-Grown Monolayer MoS2. *Current Applied Physics* **2023**, *45*, 99–104. https://doi.org/10.1016/j.cap.2022.11.008.
(81) Liu, B.; Liang, J.; Zhou, Y.; Li, L.; Li, N.; Ma, S. Space-Confined and Uniform Growth of 2D MoS2 Flakes. *Journal of Solid State Chemistry* **2024**, *332*, 124583. https://doi.org/10.1016/j.jssc.2024.124583.
(82) Pondick, J. V.; Woods, J. M.; Xing, J.; Zhou, Y.; Cha, J. J. Stepwise Sulfurization from $MoO_3$ to $MoS_2$ via Chemical Vapor Deposition. *ACS Appl. Nano Mater.* **2018**, *1* (10), 5655–5661. https://doi.org/10.1021/acsanm.8b01266.
(83) Chen, F.; Su, W. The Effect of the Experimental Parameters on the Growth of $MoS_2$ Flakes. *CrystEngComm* **2018**, *20* (33), 4823–4830. https://doi.org/10.1039/C8CE00733K.
(84) Sirat, M. S.; Johari, M. H.; Mohmad, A. R.; Haniff, M. A. S. M.; Ani, M. H.; Syono, M. I.; Mohamed, M. A. Uniform Growth of MoS2 Films Using Ultra-Low MoO3 Precursor in One-Step Heating Chemical Vapor Deposition. *Thin Solid Films* **2022**, *744*, 139092. https://doi.org/10.1016/j.tsf.2022.139092.
(85) Coordination for the Improvement of Higher Education Personnel - CAPES. *PedraQualis List*. https://predaqualis.netlify.app/.
(86) *Predatory Journals*. https://www.predatoryjournals.org/.
(87) Grasseschi, D. Systematic Review of Nanomaterials Synthesis - AuNP and MoS2. **2026**. https://doi.org/10.17605/OSF.IO/YSPHK.
(88) Grasseschi, D.; dos Santos, D. NANOMATERIAIS PLASMÔNICOS: PARTE I. FUNDAMENTOS DA ESPECTROSCOPIA DE NANOPARTÍCULAS E SUA RELAÇÃO COM O EFEITO SERS. *Química Nova* **2020**, *43* (10), 1463–1481. https://doi.org/10.21577/0100-4042.20170621.
(89) Odom, T. W.; Schatz, G. C. Introduction to Plasmonics. *Chemical reviews* **2011**, *111* (6), 3667–3668. https://doi.org/10.1021/cr2001349.
(90) Maier, S. A. *Plasmonics: Fundamentals and Applications*; Springer, Ed.; Springer US: New York, NY, 2007. https://doi.org/10.1007/0-387-37825-1.
(91) Santos, D.; Grasseschi, D. NANOMATERIAIS PLASMÔNICOS: PARTE II. QUÍMICA DE COORDENAÇÃO DE SUPERFÍCIE E SUA APLICAÇÃO EM SENSORES E CATALISADORES. *Química Nova* **2020**, *43* (10), 1482–1499. https://doi.org/10.21577/0100-4042.20170629.
(92) Xiao, M.; Jiang, R.; Wang, F.; Fang, C.; Wang, J.; Yu, J. C. Plasmon-Enhanced Chemical Reactions. *Journal of Materials Chemistry A* **2013**, *1* (19), 5790. https://doi.org/10.1039/c3ta01450a.
(93) Fonsaca, J. E. S.; Moreira, M. P.; Farooq, S.; de Araujo, R. E.; de Matos, C. J. S.; Grasseschi, D. Surface Plasmon Resonance Platforms for Chemical and Bio Sensing. In *Reference Module in Biomedical Sciences*; Elsevier, 2021. https://doi.org/10.1016/B978-0-12-822548-6.00036-4.
(94) Dykman, L.; Khlebtsov, B.; Khlebtsov, N. Drug Delivery Using Gold Nanoparticles. *Advanced Drug Delivery Reviews* **2025**, *216*, 115481. https://doi.org/10.1016/j.addr.2024.115481.
(95) Giannini, V.; Fernández-Domínguez, A. I.; Heck, S. C.; Maier, S. A. Plasmonic Nanoantennas:

Fundamentals and Their Use in Controlling the Radiative Properties of Nanoemitters. *Chemical reviews* **2011**, *111* (6), 3888–3912. https://doi.org/10.1021/cr1002672.
(96) Atwater, H. A.; Polman, A. Plasmonics for Improved Photovoltaic Devices. *Nature materials* **2010**, *9* (3), 205–213. https://doi.org/10.1038/nmat2629.
(97) Noginov, M. A.; Zhu, G.; Belgrave, A. M.; Bakker, R.; Shalaev, V. M.; Narimanov, E. E.; Stout, S.; Herz, E.; Suteewong, T.; Wiesner, U. Demonstration of a Spaser-Based Nanolaser. *Nature* **2009**, *460* (7259), 1110–1112. https://doi.org/10.1038/nature08318.
(98) Faraday, M. The Bakerian Lecture: Experimental Relations of Gold (and Other Metals) to Light. *Philosophical Transactions of the Royal Society of London* **1857**, *147*, 145–181. https://doi.org/10.1098/rstl.1857.0011.
(99) Turkevich, J.; Stevenson, P. C. P. P. C.; Hillier, J. A Study of the Nucleation and Growth Processes in the Synthesis of Colloidal Gold. *Discussions of the Faraday Society* **1951**, *11*, 55–75.
(100) Turkevich, J.; Stevenson, P. C.; Hillier, J. The Formation of Colloidal Gold. *The Journal of Physical Chemistry* **1953**, *57* (7), 670–673. https://doi.org/10.1021/j150508a015.
(101) Turkevich, J.; Garton, G.; Stevenson, P. C. The Color of Colloidal Gold. *Journal of Colloid Science* **1954**, *9*, 26–35. https://doi.org/10.1016/0095-8522(54)90070-7.
(102) Turkevich, J. Colloidal Gold. Part I. *Gold Bulletin* **1985**, *18* (3), 86–91. https://doi.org/10.1007/BF03214690.
(103) Turkevich, J. Colloidal Gold. Part II. *Gold Bulletin* **1985**, *18* (4), 125–131. https://doi.org/10.1007/BF03214694.
(104) Frens, G. Controlled Nucleation for the Regulation of the Particle Size in Monodisperse Gold Suspensions. *Nature Physical Science* **1973**, *241* (105), 20–22. https://doi.org/10.1038/physci241020a0.
(105) Chow, M. K. K.; Zukoski, C. F. F. Gold Sol Formation Mechanisms - Role of Colloidal Stability. *Journal of Colloid and Interface Science* **1994**, *165* (1), 97–109. https://doi.org/10.1006/jcis.1994.1210.
(106) Ojea-Jiménez, I.; Bastús, N. G.; Puntes, V. Influence of the Sequence of the Reagents Addition in the Citrate-Mediated Synthesis of Gold Nanoparticles. *J. Phys. Chem. C* **2011**, *115* (32), 15752–15757. https://doi.org/10.1021/jp2017242.
(107) Li, C.; Li, D.; Wan, G.; Xu, J.; Hou, W. Facile Synthesis of Concentrated Gold Nanoparticles with Low Size-Distribution in Water: Temperature and pH Controls. *Nanoscale research letters* **2011**, *6* (1), 440. https://doi.org/10.1186/1556-276X-6-440.
(108) Volkert, A. a; Subramaniam, V.; Haes, A. J. Implications of Citrate Concentration during the Seeded Growth Synthesis of Gold Nanoparticles. *Chem. Commun.* **2011**, *47* (1), 478–480. https://doi.org/10.1039/C0CC02075C.
(109) Ojea-Jiménez, I.; Bastús, N. G.; Puntes, V. Influence of the Sequence of the Reagents Addition in the Citrate-Mediated Synthesis of Gold Nanoparticles. *The Journal of Physical Chemistry C* **2011**, *115* (32), 15752–15757. https://doi.org/10.1021/jp2017242.
(110) Bastús, N. G.; Comenge, J.; Puntes, V. Kinetically Controlled Seeded Growth Synthesis of Citrate-Stabilized Gold Nanoparticles of up to 200 Nm: Size Focusing versus Ostwald Ripening. *Langmuir : the ACS journal of surfaces and colloids* **2011**, *27* (17), 11098–11105. https://doi.org/10.1021/la201938u.
(111) Schulz, F.; Homolka, T.; Bastús, N. G.; Puntes, V.; Weller, H.; Vossmeyer, T. Little Adjustments Significantly Improve the Turkevich Synthesis of Gold Nanoparticles. *Langmuir : the ACS journal of surfaces and colloids* **2014**, *30* (35), 10779–10784. https://doi.org/10.1021/la503209b.
(112) Polte, J.; Ahner, T. T.; Delissen, F.; Sokolov, S.; Emmerling, F.; Thünemann, A. F.; Kraehnert, R. Mechanism of Gold Nanoparticle Formation in the Classical Citrate Synthesis Method Derived from Coupled in Situ XANES and SAXS Evaluation. *Journal of the American Chemical Society* **2010**, *132* (4), 1296–1301. https://doi.org/10.1021/ja906506j.
(113) Thanh, N. T. K.; Maclean, N.; Mahiddine, S. Mechanisms of Nucleation and Growth of Nanoparticles in Solution. *Chemical Reviews* **2014**, *114* (15), 7610–7630. https://doi.org/10.1021/cr400544s.
(114) Kimling, J.; Maier, M.; Okenve, B.; Kotaidis, V.; Ballot, H.; Plech, A. Turkevich Method for Gold Nanoparticle Synthesis Revisited. *The Journal of Physical Chemistry B* **2006**, *110* (32),

15700–15707. https://doi.org/10.1021/jp061667w.
(115) Chaves, A.; Azadani, J. G.; Alsalman, H.; Da Costa, D. R.; Frisenda, R.; Chaves, A. J.; Song, S. H.; Kim, Y. D.; He, D.; Zhou, J.; Castellanos-Gomez, A.; Peeters, F. M.; Liu, Z.; Hinkle, C. L.; Oh, S.-H.; Ye, P. D.; Koester, S. J.; Lee, Y. H.; Avouris, Ph.; Wang, X.; Low, T. Bandgap Engineering of Two-Dimensional Semiconductor Materials. *npj 2D Mater Appl* **2020**, *4* (1), 29. https://doi.org/10.1038/s41699-020-00162-4.
(116) Yang, R.; Fan, Y.; Zhang, Y.; Mei, L.; Zhu, R.; Qin, J.; Hu, J.; Chen, Z.; Hau Ng, Y.; Voiry, D.; Li, S.; Lu, Q.; Wang, Q.; Yu, J. C.; Zeng, Z. 2D Transition Metal Dichalcogenides for Photocatalysis. *Angew Chem Int Ed* **2023**, *62* (13), e202218016. https://doi.org/10.1002/anie.202218016.
(117) Xu, Y.; Ge, R.; Yang, J.; Li, J.; Li, S.; Li, Y.; Zhang, J.; Feng, J.; Liu, B.; Li, W. Molybdenum Disulfide (MoS2)-Based Electrocatalysts for Hydrogen Evolution Reaction: From Mechanism to Manipulation. *Journal of Energy Chemistry* **2022**, *74*, 45–71. https://doi.org/10.1016/j.jechem.2022.06.031.
(118) Tan, S. J. R.; Sarkar, S.; Zhao, X.; Luo, X.; Luo, Y. Z.; Poh, S. M.; Abdelwahab, I.; Zhou, W.; Venkatesan, T.; Chen, W.; Quek, S. Y.; Loh, K. P. Temperature- and Phase-Dependent Phonon Renormalization in 1T′-$MoS_2$. *ACS Nano* **2018**, *12* (5), 5051–5058. https://doi.org/10.1021/acsnano.8b02649.
(119) Silva, W. C.; Kim, M.; De Oliveira, G. A.; Da Silva, L. V.; Tetteh, E. B.; Gomez, C. M. R.; Ramos, W. D. R.; Fragneaud, B.; Schuhmann, W.; Santana Santos, C.; Grasseschi, D. Investigation of Doping Effects on the Local Electrochemical Activity of Transition Metal Dichalcogenides 2D Materials. *Adv Funct Materials* **2024**, *34* (39), 2403224. https://doi.org/10.1002/adfm.202403224.
(120) Abidi, I. H.; Giridhar, S. P.; Tollerud, J. O.; Limb, J.; Waqar, M.; Mazumder, A.; Mayes, E. L.; Murdoch, B. J.; Xu, C.; Bhoriya, A.; Ranjan, A.; Ahmed, T.; Li, Y.; Davis, J. A.; Bentley, C. L.; Russo, S. P.; Gaspera, E. D.; Walia, S. Oxygen Driven Defect Engineering of Monolayer $MoS_2$ for Tunable Electronic, Optoelectronic, and Electrochemical Devices. *Adv Funct Materials* **2024**, *34* (37), 2402402. https://doi.org/10.1002/adfm.202402402.
(121) Lee, Y.; Zhang, X.; Zhang, W.; Chang, M.; Lin, C.; Chang, K.; Yu, Y.; Wang, J.; Chang, C.; Li, L.; Lin, T. Synthesis of Large-Area MoS2 Atomic Layers with Chemical Vapor Deposition. *ADVANCED MATERIALS* **2012**, *24* (17), 2320–2325. https://doi.org/10.1002/adma.201104798.
(122) Zhou, D.; Shu, H.; Hu, C.; Jiang, L.; Liang, P.; Chen, X. Unveiling the Growth Mechanism of MoS2 with Chemical Vapor Deposition: From Two-Dimensional Planar Nucleation to Self-Seeding Nucleation. *CRYSTAL GROWTH & DESIGN* **2018**, *18* (2), 1012–1019. https://doi.org/10.1021/acs.cgd.7b01486.
(123) Wang, L.; Sun, Y.; Kannan, K.; Gannon, L.; Guo, X.; Rafferty, A.; Gaff, K.; Mullani, N. B.; Weng, H.; Zhou, Y.; Nicolosi, V.; McGuinness, C.; Gleeson, P.; Zhang, H. Enhanced Chemical Vapor Deposition of Monolayer $MoS_2$ Films via a Clean Promoter. *J. Phys.: Condens. Matter* **2025**, *37* (46), 465001. https://doi.org/10.1088/1361-648X/ae14c8.
(124) Zafar, A.; Zafar, Z.; Zhao, W.; Jiang, J.; Zhang, Y.; Chen, Y.; Lu, J.; Ni, Z. Sulfur-Mastery: Precise Synthesis of 2D Transition Metal Dichalcogenides. *ADVANCED FUNCTIONAL MATERIALS* **2019**, *29* (27). https://doi.org/10.1002/adfm.201809261.
(125) Heiserer, S.; Eder, P.; Coileáin, C.; Biba, J.; Stimpel-Lindner, T.; Bartlam, C.; Rührmair, U.; Duesberg, G. Controllable and Reproducible Growth of Transition Metal Dichalcogenides by Design of Experiments. *ADVANCED ELECTRONIC MATERIALS* **2023**, *9* (10). https://doi.org/10.1002/aelm.202300281.
(126) Withanage, S.; Kalita, H.; Chung, H.; Roy, T.; Jung, Y.; Khondaker, S. Uniform Vapor-Pressure-Based Chemical Vapor Deposition Growth of MoS2 Using MoO3 Thin Film as a Precursor for Coevaporation. *ACS OMEGA* **2018**, *3* (12), 18943–18949. https://doi.org/10.1021/acsomega.8b02978.
(127) Yang, S.; Shim, G.; Seo, S.; Choi, S. Effective Shape-Controlled Growth of Monolayer MoS2 Flakes by Powder-Based Chemical Vapor Deposition. *NANO RESEARCH* **2017**, *10* (1), 255–262. https://doi.org/10.1007/s12274-016-1284-6.
(128) Narayanan, V.; Gokul, M.; Rahman, A. How to ?Train? Your CVD to Grow Large-Area 2D Materials. *MATERIALS RESEARCH EXPRESS* **2019**, *6* (12). https://doi.org/10.1088/2053-

1591/ab5383.
(129) Wei, J.; Huang, J.; Du, J.; Bian, B.; Li, S.; Wang, D. Effect of the Geometry of Precursor Crucibles on the Growth of MoS2 Flakes by Chemical Vapor Deposition. *NEW JOURNAL OF CHEMISTRY* **2020**, *44* (48), 21076–21084. https://doi.org/10.1039/d0nj05486k.
(130) Collins, H. M. Tacit Knowledge, Trust and the Q of Sapphire. *Soc Stud Sci* **2001**, *31* (1), 71–85. https://doi.org/10.1177/030631201031001004.
(131) Bøggild, P.; Booth, T. J.; Jessen, B. S.; Shivayogimath, A.; Lassaline, N.; Hofmann, S.; Daasbjerg, K.; Smith, A.; Nørgaard, K.; Zurutuza, A.; Barkan, T.; Pollard, A. J. Recommendations and Tools to Enable Reproducibility in 2D Materials Research. arXiv October 9, 2025. https://doi.org/10.48550/arXiv.2409.18994.
(132) Bøggild, P. Research on Scalable Graphene Faces a Reproducibility Gap. *Nat Commun* **2023**, *14* (1), 1126. https://doi.org/10.1038/s41467-023-36891-5.
(133) Bøggild, P.; Booth, T. J.; Lassaline, N.; Jessen, B. S.; Hofmann, S.; Daasbjerg, K.; Smith, A.; Nørgaard, K.; Zurutuza, A.; Asselberghs, I.; Barkan, T.; Taboryski, R. Closing the Reproducibility Gap: 2D Materials Research.